\documentclass[prd, preprint, amsmath, amssymb, aps, showpacs, superscriptaddress]{revtex4-1}
\usepackage{mathptmx}
\usepackage{helvet}
\usepackage{amssymb}
\usepackage{amsmath}
\usepackage{graphicx}
\usepackage{amsthm}
\usepackage{color}
\usepackage{ulem}
\numberwithin{equation}{section}
\theoremstyle{plain}

  \theoremstyle{plain}
  
  \theoremstyle{plain}

\begin{document}
\title{Dipole Modulation of Cosmic Microwave Background Temperature and Polarization.}
\author{Shamik Ghosh}
\email{shamik@iitk.ac.in}
\author{Rahul Kothari}
\email{rahulko@iitk.ac.in}
\author{Pankaj Jain}
\email{pkjain@iitk.ac.in}
\affiliation{Dept. of Physics, Indian Institute of Technology,\\
Kanpur - 208016, India}
\author{Pranati K. Rath}
\email{pranati@iopb.res.in}
\affiliation{Institute  of Physics, \\
Sachivalaya Marg, Bhubaneswar - 751005, Odisha, India}

\date{\today}

\begin{abstract}
We propose a dipole modulation model for 
the Cosmic Microwave Background Radiation (CMBR) polarization field. 
We show that the model leads to correlations between $l$ and $l+1$ 
multipoles, exactly as in the case of temperature. We obtain 
results for the case of $TE$, $EE$ and $BB$ correlations. 
An anisotropic or inhomogeneous model of primordial power spectrum which
leads to such correlations in temperature field also predicts similar
correlations in 
 CMBR polarization. 
We analyze the CMBR temperature
and polarization data in order to extract the signal of 
these correlation between $l$ and $l+1$ multipoles.  
 Our results for
the case of temperature using the latest PLANCK data agree with those
obtained by an earlier analysis. 
A detailed study of the correlation in the polarization data is not 
possible at present. 
Hence we restrict ourselves to a preliminary investigation in this case.

\end{abstract}


\maketitle

\maketitle
\flushbottom 
\section{Introduction}
The Cosmic Microwave Background Radiation (CMB) shows a hemispherical 
power asymmetry, due to which the power in the two hemispheres is significantly 
different \cite{Eriksen2004,Eriksen2007,Erickcek2008,Hansen2009,Hoftuft2009,Paci2013,Planck2013a,Akrami2014}. 
A dipole modulation of a statistically isotropic signal provides a useful
parametrization of the observed power asymmetry. According to the model, 
the observed temperature fluctuation $\Delta \tilde{T}$ along a direction $\hat{n}$ is expressed as \cite{Gordon2005,Gordon2007,Prunet2005,Bennett2011},
\begin{equation}
\Delta \tilde{T}\left(\hat{n}\right) = \Delta T\left(\hat{n}\right)\left(1+A\hat{\lambda}_1\cdot\hat{n}\right),
\label{eq:model}
\end{equation}
where $\Delta T\left(\hat{n}\right)$ is a statistically isotropic field, $A$ 
 the dipole amplitude and $\hat{\lambda}_1$ the dipole direction. 
Throughout this paper we shall denote the observed fields, which are 
assumed to have some contribution due to dipole modulation, with a tilde
and the corresponding fields in an isotropic model without a tilde.
Choosing our axes such that $\hat{\lambda}_1$ is along $\hat{z}$, Eq. (\ref{eq:model}) can be written as
\begin{equation}
\Delta \tilde{T}\left(\hat{n}\right) = \Delta T\left(\hat{n}\right)\left(1+A \cos \theta\right).
\label{modelz}
\end{equation}
As shown in \cite{Prunet2005,Rath:2013}, the two point correlation of such a modulated temperature field would show correlations between $l$ and $l + 1$.

If the observed signal of hemispherical anisotropy or equivalently
dipole modulation is related to a physical effect, 
 we expect a similar signal to be present in polarization fields too.
Several studies have associated this effect with a primordial 
inhomogeneous or anisotropic model. Such models lead to a modification
of the primordial power spectrum which culminates in depicting correlations between 
different multipoles similar to those predicted by Eq. \ref{eq:model}. 
In a recent paper it has been shown that such a primordial model 
also leads to correlations between $l$ and $l + 1$ of the polarization
fields.  
In this paper we propose a dipole modulation model for the CMBR polarization 
field, analogous to Eq. \ref{eq:model}. 
Such a model is useful to empirically characterize the observed hemispherical
anisotropy that might be present in the polarization data, irrespective
of the physical cause of its origin.
We show that this model leads to correlations between the $l$ and $l+1$
multipoles for the polarization fields. We also determine the explicit
form of these correlations. 

We search for such correlations in the recently released Planck experiment data in both temperature and polarization signals and compare them with previous results and predictions. In the case of polarization a detailed study is not 
possible due to difficulty in handling and interpretation of the noise
files. Hence in this case we restrict ourselves to a preliminary investigation.

\section{Test for dipole modulation}
The modulated temperature field is given by Eq. (\ref{modelz}). 
This being a field on a sphere, can be expanded in spherical harmonics as
\begin{equation}
\Delta \tilde{T} \left(\hat{n}\right)=\sum_{l,m} \tilde {a}^T_{lm} Y_{lm}\left(\hat{n}\right)
\end{equation}
The two point correlation of the temperature field in multipole space can be written as 

\begin{equation}
\langle \tilde{a}^T_{lm}\tilde{a}^{T*}_{l'm'}\rangle =%(-1)^{m+m'}%
\int d \Omega_{\hat{n}}d \Omega_{\hat{n}'}Y^{*}_{lm}\left(\hat{n}\right)Y_{l'm'}\left(\hat{n}'\right) \langle\Delta \tilde{T}\left(n\right)\Delta \tilde{T}\left(\hat{n}'\right)\rangle
\label{correlation}
\end{equation}
As shown in \cite{Rath:2013}, using Eq. (\ref{modelz}) we obtain
\begin{equation}
\langle \tilde{a}^T_{lm} \tilde{a}^{T*}_{l'm'}\rangle = C^T_l \delta_{ll'}\delta_{mm'}+A \left(C^T_{l'}+C^T_l\right)\xi^0_{lm;l'm'}
\label{corr_TT}
\end{equation}
where 
\begin{equation}
\xi^0_{lm;l'm'}=\delta_{mm'}\left[\sqrt{\frac{(l+m+1)(l-m+1)}{(2l+3)(2l+1)}}\delta_{l',l+1}+\sqrt{\frac{(l+m)(l-m)}{(2l+1)(2l-1)}}\delta_{l',l-1}\right]
\end{equation} 
In Eq. (\ref{corr_TT}), the first term on RHS corresponds to  the isotropic part of the correlation $\langle \tilde{a}^T_{lm} \tilde{a}^{T*}_{l'm'}\rangle_{\text{iso}}$ and the second term is the contribution of the modulation, 
$\langle \tilde{a}^T_{lm} \tilde{a}^{T*}_{l'm'}\rangle_{\text{mod}}$.
As we shall see the multipole power $C^T_l$ does not get any contribution
from the modulation term. 
This is found to be true also for the power in the polarization fields,
to be discussed later.
Hence we do not denote it with a tilde. 
We follow \cite{Rath:2013} and seek correlations between $l$ and $l + 1$
multipoles, which can be expressed as,
\begin{equation}
\langle \tilde{a}^T_{lm} \tilde{a}^{T*}_{l+1m}\rangle = A\left[C^T_{l+1}+C^T_l\right]\sqrt{\frac{(l+m+1)(l-m+1)}{(2l+3)(2l+1)}}\,.
\end{equation}

Theoretical models used to explain the dipole modulation of the temperature field 
predict a similar correlation between $l$ and $l+1$ multipoles 
 in the CMB E-mode polarization field \cite{Namjoo:2014pqa,Kothari:2015} 
in the same direction. 
These predictions may be tested in future 
 by determining these correlations in the CMB polarization field.

A detailed discussion of CMB polarization is contained in \cite{Zaldarriaga:1996,Kamionkowski:1996} and here we use the notation of \cite{Zaldarriaga:1996}. 
The CMB polarization field is characterized by two Stokes parameters 
$\tilde Q$ and $\tilde U$, while the temperature 
fluctuation field corresponds to Stokes' parameter $\tilde I$. 
Here $\tilde Q$ and $\tilde U$ denote the
dipole modulated polarization fields.  
Under a rotation by an 
angle $\psi$, the temperature field transforms as a scalar, 
while combinations of 
$\tilde Q$ and $\tilde U$ behave as spin $\pm2$ fields on a sphere, viz.
\begin{equation}
(\tilde Q\pm i\tilde U)'(\hat{n}) = e^{\mp 2 i \psi}(\tilde Q\pm i\tilde U)(\hat{n}),
\end{equation} 
and can be expanded in spin $\pm2$ harmonics as
\begin{equation}
(\tilde Q\pm i\tilde U)(\hat{n})=\sum_{lm}\tilde a_{\pm2,lm\ \pm 2}Y_{lm}(\hat{n}).
\label{spin2}
\end{equation}
Using the spin raising and lowering operators $\eth$ and $\bar{\eth}$, spin 0 objects can be constructed from $\tilde Q$ and $\tilde U$ fields \cite{Newman:1966}. Using $\eth$ and $\bar{\eth}$ suitably on Eq. (\ref{spin2}) we get
 \begin{align}
 \bar{\eth}^2(\tilde Q+ i\tilde U)(\hat{n})=&\sum_{lm}\sqrt{\frac{(l+2)!}{(l-2)!}}\tilde a_{2,lm}\ Y_{lm}(\hat{n})\\
 \eth^2(\tilde Q- i\tilde U)(\hat{n})=&\sum_{lm}\sqrt{\frac{(l+2)!}{(l-2)!}}\tilde a_{-2,lm}\ Y_{lm}(\hat{n})\,.
 \end{align}
 Finally the standard $E$ and $B$ mode polarization field can be expressed as,
\begin{align}
\widetilde{E}(\hat{n})=&-\frac{1}{2}\left[\bar{\eth}^2(\tilde Q+i\tilde U)+
\eth^2(\tilde Q-i\tilde U)\right]=\sum_{lm}\sqrt{\frac{(l+2)!}{(l-2)!}}\tilde a^E_{lm}Y_{lm}(\hat{n}) \label{Emode}\\
\widetilde{B}(\hat{n})=&\frac{i}{2}\left[\bar{\eth}^2(\tilde Q+i\tilde U)-
\eth^2(\tilde Q-i\tilde U)\right]=\sum_{lm}\sqrt{\frac{(l+2)!}{(l-2)!}}\tilde a^B_{lm}Y_{lm}(\hat{n})\,. \label{Bmode}
\end{align}
The coefficients $\tilde a^E_{lm}$ and $\tilde a^B_{lm}$ are defined as linear combinations of $\tilde a_{\pm2,lm}$ as:
\begin{align}
 \tilde a^E_{lm}=&-\frac{1}{2}\left(\tilde a_{2,lm}+\tilde a_{-2,lm}\right)\\
 \tilde a^B_{lm}=&\frac{i}{2}\left(\tilde a_{2,lm}-\tilde a_{-2,lm}\right)
\label{eq:aEaB}
\end{align}
The $\tilde a^E_{lm}$s define the E-mode polarization in multipole space and are unchanged under parity transformation in contrast to $\tilde a^B_{lm}$s which do change sign under such a transformation. 
 The scalar fields defined in Eqs. (\ref{Emode}) and (\ref{Bmode}) are 
the real space constructs of the E-mode and B-mode polarizations representing
 the irrotational and curl components of the CMB polarizations respectively. 
In this work we are interested in the E-mode field. We define the auto correlation of the E field and cross correlation of the E and T fields as
\begin{align}
C_l^{EE} =&\frac{1}{2l+1}\sum_m\langle \tilde a^E_{lm}\tilde a^{E*}_{lm}\rangle\\
C_l^{TE} =&\frac{1}{2l+1}\sum_m\langle \tilde a^E_{lm}\tilde a^{T*}_{lm}\rangle
\end{align}

In order to study the $l$ and $l+1$ correlations we construct 
\begin{equation}
C^{XX}_{l,l+1} = \frac{l(l+1)}{(2l+1)}\sum_{m=-l}^{m=+l} \langle\tilde a^X_{lm}\tilde a^{X*}_{l+1m}\rangle,
\label{eq:Cll1XX}
\end{equation}
and define our statistics as 
\begin{equation}
S^{XX}_H = \sum_{l_{\text{min}}}^{l_{\text{max}}} C^{XX}_{l,l+1}.
\end{equation}
Here $X$ can be either $T$ or $E$ giving us $TT$, $EE$, $TE$ and $ET$ correlations.
We search the direction for which the statistic $S^{XX}_H$ maximizes in each of the maps. We also define a quantity $R$ as the ratio of the anisotropic part to isotropic part, i.e., 
\begin{equation}
R = \frac{\sum_{l_{\text{min}}}^{l_{\text{max}}}C^{XX}_{l,l+1}}{\sum_{l_{\text{min}}}^{l_{\text{max}}}l(l+1)C^{XX}_l}
\end{equation}
This may be seen as a measure of the fraction of the anisotropic effect to the isotropic power.

\section{Dipole Modulation in Polarization}

The 
dipole modulated polarization fields are denoted by 
 $\tilde{Q}\left(\hat{n}\right)$ and $\tilde{U}\left(\hat{n}\right)$
where $\hat{n}\equiv\left(\theta,\phi\right)$. 
We also define
\begin{equation}
\tilde{\alpha}_\pm\left(\hat{n}\right)  = 
\tilde{Q}\left(\hat{n}\right)\pm i\tilde{U}\left(\hat{n}\right)\,,
\end{equation}
$\alpha_\pm(\hat n)=Q(\hat n)\pm iU(\hat n)$ where $Q$ and $U$ are the standard unmodulated
fields in an isotropic model,  $\tilde\alpha_-=\tilde\alpha_+^*$
and $\alpha_-=\alpha_+^*$. 
The preferred direction $\hat \lambda$
is taken to be the same for
both $\tilde Q$ and $\tilde U$ as well as the temperature field \cite{Kothari:2015}. 
In analogy with temperature, we propose the following model for 
dipole modulation of polarization:
\begin{eqnarray}
\tilde{\alpha}_+\left(\hat{n}\right) & = &
\alpha_+\left(\hat{n}\right)\left(1+A_P\hat\lambda\cdot\hat{n}\right),\nonumber\\
\tilde{\alpha}_-\left(\hat{n}\right) & = &
\alpha_-\left(\hat{n}\right)\left(1+A_P^*\hat\lambda\cdot\hat{n}\right)\,.
\label{eq:alpha_modu}
\end{eqnarray}
Here $A_P=A_1+iA_2$ is a complex parameter. 
We choose our coordinates such that $\hat\lambda=\hat z$ and hence
 $\hat\lambda\cdot\hat{n}=\cos\theta$.
In terms of the Stokes' parameters, we obtain
\begin{eqnarray}
\tilde{Q} & = & Q(1+A_1\cos\theta)-UA_2\cos\theta
\nonumber\\
\tilde{U} & = & QA_2\cos\theta+U(1+A_1\cos\theta)
\label{eq:QUmod}
\end{eqnarray}
Using Eqs. \ref{spin2} and \ref{eq:aEaB}
for the modulated polarization fields, we obtain
\begin{equation}
\tilde{\alpha}_\pm  = -\sum_{lm}\left(\tilde{a}^{\,E}_{lm}\pm 
i\tilde{a}^{\,B}_{lm}\right) \ {}_{\pm2}Y_{lm}\,,
\end{equation}
where $\tilde{a}_{E,lm}$ and $\tilde{a}_{B,lm}$ denote the harmonic 
coefficients of the modulated fields. 
Inverting the above equation we obtain
\begin{equation}
-\left(\tilde{a}^{\,E}_{lm}\pm i\tilde{a}^{\,B}_{lm}\right)  = 
\int\tilde{\alpha}_\pm\left(\hat{n}\right)\ {}_{\pm 2}Y_{lm}^*(\hat{n})
d\Omega\,.
\end{equation}
This leads to
\begin{equation}
\tilde{a}^{\,E}_{lm}=-\frac{1}{2}\int\left[\alpha_+\left(\hat{n}\right)
(1+A_P\cos\theta) {}_2Y_{lm}^*(\hat{n})+\alpha_-(\hat{n})(1+A_P^*
\cos\theta) {}_{-2}Y_{lm}^*(\hat{n})\right]d\Omega,\label{eq:rep_alm_E}
\end{equation}
where we have used Eq. (\ref{eq:alpha_modu}). We also obtain a similar 
equation for $\tilde{a}_{B,lm}$.

\subsection{\label{sec:correlations} Correlations of the Dipole 
Modulated Polarization Field}

The two point correlations of the dipole modulated $E$ field
harmonic coefficients can be expressed as,  
\begin{equation}
\left\langle \tilde{a}^{\,E}_{lm}\tilde{a}_{l^{\prime}m^{\prime}}^{\,E*}
\right\rangle ={1\over 4} (I_1+I_2+I_3+I_4)
\label{eq:atildeE2}
\end{equation}
where
\begin{eqnarray*}
I_1 &=& \iint d\Omega d\Omega^{\prime}
\left\langle 
\alpha_+(\hat{n})\alpha_-(\hat{n}^{\prime})\right\rangle
\left(1+A_P\cos\theta\right)\left(1+A_P^*\cos\theta^{\prime}\right)
{}_2Y_{lm}^*(\hat{n})\  {}_2Y_{l^{\prime}m^{\prime}}
(\hat{n}^{\prime})
,\\
I_2 &=& \iint
d\Omega d\Omega^{\prime}
\left\langle
\alpha_+(\hat{n})\alpha_+(\hat{n}^{\prime})\right\rangle
\left(1+A_P\cos\theta\right)\left(1+A_P\cos\theta^{\prime}\right)
{}_2Y_{lm}^*(\hat{n})\ {}_{-2}Y_{l^{\prime}m^{\prime}}
(\hat{n}^{\prime}),\\
I_3 & = & \iint
d\Omega d\Omega^{\prime}
\left\langle
\alpha_-(\hat{n})\alpha_-(\hat{n}^{\prime})\right\rangle
\left(1+A_P^*\cos\theta\right)\left(1+A_P^*\cos\theta^{\prime}\right)
{}_{-2}Y_{lm}^*(\hat{n})
\ {}_2Y_{l^{\prime}m^{\prime}}(\hat{n}^{\prime}),\\
I_{4} & = & \iint d\Omega d\Omega^{\prime} \left\langle
\alpha_-(\hat{n})\alpha_+(\hat{n}^{\prime})\right\rangle
\left(1+A_P^*\cos\theta\right)\left(1+A_P\cos\theta^{\prime}\right)
{}_{-2}Y_{lm}^*(\hat{n})\ {}_{-2}Y_{l^{\prime}m^{\prime}}
(\hat{n}^{\prime}).
\end{eqnarray*}
The two point correlations appearing on the right hand side of these equations
can written as:
\begin{eqnarray*}
\left\langle
\alpha_+(\hat{n})\alpha_-(\hat{n}^{\prime})\right\rangle & = &
\sum_{l^{\prime\prime}m^{\prime\prime}}\left(C^{EE}_{l^{\prime\prime}}+C^{BB}_{l^{\prime\prime}}\right)
{}_2Y_{l^{\prime\prime}m^{\prime\prime}}(\hat{n})
\ {}_2Y_{l^{\prime\prime}m^{\prime\prime}}^*(\hat{n}^{\prime}),\\
\left\langle
\alpha_+(\hat{n})\alpha_+(\hat{n}^{\prime})\right\rangle & = &
\sum_{l^{\prime\prime}m^{\prime\prime}}\left(C^{EE}_{l^{\prime\prime}}-
C^{BB}_{l^{\prime\prime}}\right)
\left(-1\right)^{m^{\prime\prime}}
{}_2Y_{l^{\prime\prime}m^{\prime\prime}}(\hat{n})
\ {}_2Y_{l^{\prime\prime}\left(-m^{\prime\prime}\right)}(\hat{n}^{\prime}),\\
\left\langle
\alpha_-(\hat{n})\alpha_-(\hat{n}^{\prime})\right\rangle & = &
\sum_{l^{\prime\prime}m^{\prime\prime}}\left(C^{EE}_{l^{\prime\prime}}-
C^{BB}_{l^{\prime\prime}}\right)\left(-1\right)^{m^{\prime\prime}}
{}_2Y_{l^{\prime\prime}m^{\prime\prime}}^*(\hat{n}^{\prime})\ {}_2Y_{l^{\prime\prime}\left(-m^{\prime\prime}\right)}^*(\hat{n}),\\
\left\langle \alpha_-(\hat{n})\alpha_+\hat{n}^{\prime})\right\rangle
& = &
\sum_{l^{\prime\prime}m^{\prime\prime}}\left(C^{EE}_{l^{\prime\prime}}+
C^{BB}_{l^{\prime\prime}}\right)
{}_2Y_{l^{\prime\prime}\left(-m^{\prime\prime}\right)}^*(\hat{n})
\ {}_2Y_{l^{\prime\prime}\left(-m^{\prime\prime}\right)}(\hat{n}^{\prime}).
\end{eqnarray*}
where we have used,
$\left\langle a^{\,E}_{lm}a_{l^{\prime}m^{\prime}}^{\,E*}\right\rangle
=C_{l}^{E}\delta_{ll^{\prime}}\delta_{mm^{\prime}}$,
$\left\langle a^{\,B}_{lm}a_{B,l^{\prime}m^{\prime}}^{\,B*}\right\rangle
=C_{l}^{B}\delta_{ll^{\prime}}\delta_{mm^{\prime}}$, 
$\left\langle a^{\,E}_{lm}a_{l^{\prime}m^{\prime}}^{\,B*}\right\rangle =0$ and
${}_{-2} Y_{lm}^*=\left(-1\right)^{m}{}_2Y_{l\left(-m\right)}$.
Here $C_{l}^{E}$ and
$C_{l}^{B}$ represent the isotropic power spectrum corresponding to 
$E$ or $B$ modes respectively. As we shall see the anisotropic model does 
not contribute to the power spectrum. Hence these also represent the 
power of the tilde fields. 

Substituting the resulting expressions of $I_i$ in Eq. \ref{eq:atildeE2},
we obtain
\begin{equation}
\left\langle \tilde{a}^{\,E}_{lm}\tilde{a}_{l^{\prime}m^{\prime}}^{\,E*}
\right\rangle = 
C_{l}^{EE}\delta_{ll^{\prime}}\delta_{mm^{\prime}} 
+ {1\over 4} \left(M_1+M_2+M_3+M_4\right)
\label{eq:atildeE21}
\end{equation}
where $M_i$ represent the corrections due to dipole modulation and
are given by,
\begin{eqnarray*}
M_{1} & = &
\sum_{l^{\prime\prime}m^{\prime\prime}}\left(C^{EE}_{l^{\prime\prime}}+
C^{BB}_{l^{\prime\prime}}\right)\iint
d\Omega d\Omega^{\prime}
\left(A_P\cos\theta+A_P^*\cos\theta^{\prime}\right) {}_2Y_{l^{\prime\prime}m^{\prime\prime}}(\hat{n})
\ {}_2Y_{l^{\prime\prime}m^{\prime\prime}}^*(\hat{n}^{\prime})
\ {}_2Y_{lm}^*(\hat{n}) \ {}_2Y_{l^{\prime}m^{\prime}}(\hat{n}^{\prime})
,\\
M_{2} & = &
\sum_{l^{\prime\prime}m^{\prime\prime}}\left(C^{EE}_{l^{\prime\prime}}-
C^{BB}_{l^{\prime\prime}}\right)\left(-1\right)^{m^{\prime\prime}+m^{\prime}}
\iint
d\Omega d\Omega^{\prime}
\left(A_P\cos\theta^{\prime}+A_P\cos\theta\right){}_2Y_{l^{\prime\prime}m^{\prime\prime}}(\hat{n})
\ {}_2Y_{l^{\prime\prime}\left(-m^{\prime\prime}\right)}(\hat{n}^{\prime})\\
& &\ \ \ \ \ \ \ \ \ \  \ \ \ \ \ \ \ \ \ \ \ \ \ \ \ \ \ \ \ \ \ \ \ \ \ \ \ \ \ \ \ \ \ \ \ \ \ \ \ \ \ \ \ \ \ \ \ \ \ 
 \times\, {}_2Y_{lm}^*(\hat{n})
\ {}_2Y_{l^{\prime}\left(-m^{\prime}\right)}^*
(\hat{n}^{\prime})
,\\
M_{3} & = &
\sum_{l^{\prime\prime}m^{\prime\prime}}\left(C^{EE}_{l^{\prime\prime}}-
C^{BB}_{l^{\prime\prime}}\right)\left(-1\right)^{m^{\prime\prime}+m}
\iint
d\Omega d\Omega^{\prime}
\left(A_P^*\cos\theta+A_P^*\cos\theta^{\prime}\right)
{}_2Y_{l^{\prime\prime}m^{\prime\prime}}^*(\hat{n}^{\prime})
\ {}_2Y_{l^{\prime\prime}\left(-m^{\prime\prime}\right)}^*(\hat{n})\\
& &\ \ \ \ \ \ \ \ \ \  \ \ \ \ \ \ \ \ \ \ \ \ \ \ \ \ \ \ \ \ \ \ \ \ \ \ \ \ \ \ \ \ \ \ \ \ \ \ \ \ \ \ \ \ \ \ \ \ \ 
\times\, {}_2Y_{l\left(-m\right)}(\hat{n})
\ {}_2Y_{l^{\prime}m^{\prime}}(\hat{n}^{\prime})
,\\
M_{4} & = &
\sum_{l^{\prime\prime}m^{\prime\prime}}\left(C^{EE}_{l^{\prime\prime}}+
C^{BB}_{l^{\prime\prime}}\right)\left(-1\right)^{m^{\prime}+m}\iint
d\Omega d\Omega^{\prime}
\left(A_P^*\cos\theta+A_P\cos\theta^{\prime}\right)
{}_2Y_{l^{\prime\prime}\left(-m^{\prime\prime}\right)}^*(\hat{n})
\ {}_2Y_{l^{\prime\prime}\left(-m^{\prime\prime}\right)}
(\hat{n}^{\prime})\\
& &\ \ \ \ \ \ \ \ \ \  \ \ \ \ \ \ \ \ \ \ \ \ \ \ \ \ \ \ \ \ \ \ \ \ \ \ \ \ \ \ \ \ \ \ \ \ \ \ \ \ \ \ \ \ \ \ \ \ \ 
\times\, {}_2Y_{l\left(-m\right)}(\hat{n})
\ {}_2Y_{l^{\prime}\left(-m^{\prime}\right)}^*(\hat{n}^{\prime})
.
\end{eqnarray*}
where we have assumed that the modulation
parameters $A_1$ and $A_2$ are small and dropped higher order terms. 
We can evaluate these integrals by using
\begin{equation}
\int_{0}^{2\pi}\int_{0}^{\pi} {}_2Y_{lm}\left(\hat{n}\right)
\ {}_2Y_{l^{\prime}m^{\prime}}^*\left(\hat{n}\right)d\Omega=\delta_{ll^{\prime}}\delta_{mm^{\prime}}.
\label{eq:J_Integral}
\end{equation}
Furthermore we define
\begin{equation}
\mathbb{I}\left(l,m,l^{\prime},m^{\prime}\right)  = 
\int_{0}^{2\pi}\int_{0}^{\pi} {}_2Y_{lm}\left(\hat{n}\right)\ {}_2Y_{l^{\prime}m^{\prime}}^*\left(\hat{n}\right)\cos\theta
d\Omega=\delta_{m,m^{\prime}}\mathbb{K}\left(l,l^{\prime},m\right),
\label{eq:I_Integral}\\
\end{equation}
This integral can be expressed in terms of the Wigner 3-j symbols by using
\begin{equation}
Y_{10}(\hat {n}) = \sqrt{3\over 4\pi}\, \cos\theta\,.
\end{equation}
 We obtain
\begin{equation}
\mathbb{I}\left(l,m,l^{\prime},m^{\prime}\right)  = 
\left(-1\right)^{m^{\prime}}\sqrt{\left(2l+1\right)\left(2l^{\prime}+1\right)}\left(\begin{array}{ccc}
l^{\prime} & l & 1\\
2 & -2 & 0
\end{array}\right)
\left(\begin{array}{ccc}
l^{\prime} & l & 1\\
-m^{\prime} & m & 0
\end{array}\right)
\end{equation}

The Wigner 3-j symbol obeys the condition,
\begin{equation}
\left(\begin{array}{ccc}
l_1 & l_2 & l_3\\
m_1 & m_2 &m_3 
\end{array}\right)
=0 \ \ \  {\rm if} \ \ \  |l_1-l_2|> l_3
\end{equation}
Using this we find that 
\begin{equation}
\mathbb{K}\left(l,l^{\prime},m\right) = 0\ \ \ \ \ {\rm if}\ 
l^{\prime}>l+1 \ {\rm and}\ l^{\prime}<l-1 
\end{equation}
For the remaining cases, $l=l^{\prime}$ and $l^{\prime}=l\pm1$, it can
be expressed as, 
\begin{equation}
\mathbb{K}\left(l,l^{\prime},m\right)=\left(-1\right)^{l+l^{\prime}}
H\left(l,l^{\prime},m\right)
\mathbb{U}\left(l,l^{\prime},m\right),
\label{eq:Kllpm}
\end{equation}
where 
\[
{H}\left(l,l^{\prime},m\right)=-2\, \sqrt{\frac{\left(2l+1\right)\left(2l^{\prime}+1\right)\left(l-m\right)!\left(l+m\right)!\left(l^{\prime}-m\right)!\left(l^{\prime}+m\right)!}{\left(l+2\right)!\left(l-2\right)!\left(l^{\prime}+2\right)!\left(l^{\prime}-2\right)!}}.
\]
\[
\mathbb{U}\left(l,l^{\prime},m\right)=\frac{(2l)!(l^{\prime}+2)!
(l^{\prime}-2)!\delta_{l+1,l^{\prime}}}{(2l+3)!(l-m)!(l+m)!}+
\frac{2m(2l)!(l+2)!(l-2)!\delta_{l,l^{\prime}}}{l(l+m)!(l-m)!
(2l+2)!}
+\frac{(2l^{\prime})!(l-2)!(l+2)!\delta_{l-1,l^{\prime}}}{(l^{\prime}+m)!
(l^{\prime}-m)!(2l+1)!}.
\]
The function $\mathbb{K}\left(l,l^{\prime},m\right)$ is explicitly evaluated
in the next subsection. 
We can now write the integrals $M_i$ as
\begin{eqnarray*}
M_{1} & = &
\left(-1\right)^{l+l^{\prime}}\delta_{mm^{\prime}}H\left(l^{\prime},l,m\right)\mathbb{U}\left(l^{\prime},l,m\right)\left[A_P\left(C^{EE}_{l^{\prime}}+
C^{BB}_{l^{\prime}}\right)+A_P^*\left(C^{EE}_{l}+C^{BB}_{l}\right)\right],\\
M_{2} & = &
\left(-1\right)^{l+l^{\prime}}\delta_{mm^{\prime}}\,A_P
\left[H\left(l,l^{\prime},-m\right)\mathbb{U}\left(l,l^{\prime},-m\right)
\left(C^{EE}_{l}-C^{BB}_{l}\right)+
H\left(l^{\prime},l,m\right)\mathbb{U}\left(l^{\prime},l,m\right)
\left(C^{EE}_{l^{\prime}}-C^{BB}_{l^{\prime}}\right)
\right],\\
M_{3} & = &
\left(-1\right)^{l+l^{\prime}}\delta_{mm^{\prime}}\,A_P^*
\left[ H\left(l,l^{\prime},-m\right)\mathbb{U}\left(l,l^{\prime},-m\right)\left(C^{EE}_{l^{\prime}}-C^{BB}_{l^{\prime}}\right)+H\left(l^{\prime},l,m\right)\mathbb{U}\left(l^{\prime},l,m\right)\left(C^{EE}_{l}-C^{BB}_{l}\right)\right],\\
M_{4} & = &
\left(-1\right)^{l+l^{\prime}}\delta_{mm^{\prime}}H\left(l,l^{\prime},-m\right)\mathbb{U}\left(l,l^{\prime},-m\right)\left[A_P^*\left(C^{EE}_{l^{\prime}}+
C^{BB}_{l^{\prime}}\right)+A_P\left(C^{EE}_{l}+C^{BB}_{l}\right)\right].
\end{eqnarray*}
where we have used
$$\mathbb{U}\left(l,l^{\prime},m\right)=\mathbb{U}\left(l^{\prime},l,m\right)
$$
and
$$
{H}\left(l,l^{\prime},m\right)={H}\left(l^{\prime},l,m\right)={H}\left(l,l^{\prime},-m\right)={H}\left(l,l^{\prime},-m\right).$$
This finally leads to 
$$
\left\langle \tilde{a}^{\,E}_{lm}\tilde{a}_{l^{\prime}m^{\prime}}^{\,E*}
\right\rangle
=  C^{EE}_{l}\delta_{ll^{\prime}}\delta_{mm^{\prime}}
+\delta_{mm^{\prime}}
{1\over 2}
\left[\mathbb{K}(l,l^{\prime},m)
\left(A_PC^{EE}_{l^{\prime}}+A_P^*C^{EE}_{l}\right)+\mathbb{K}(l,l^{\prime},-m)\left(A_P^*C_{l^{\prime}}^{EE}+A_PC_{l}^{EE}\right)\right]
.
$$

The first term on the right hand side
of this equation is the standard contribution due
to an isotropic field. The second term arises due to dipole modulation. 
In this term only contributions linear in the dipole parameters $A_1$ and 
$A_2$ have been kept. 
We see that the modulation model, Eq. \ref{eq:alpha_modu},
leads to correlations between multipoles $l$ and $l+1$ besides also leading
to additional contributions proportional to $\delta_{mm'}\delta_{ll'}$.  
However the latter contributions cancel out after summing over $m$ and
can be ignored. 
Hence after summing over $m$ the dipole modulation term leads to
correlations only between $l$ and $l+1$. 
We also notice that the terms proportional to $\delta_{l+1,l'}$
and $\delta_{l-1,l'}$ in Eq. \ref{eq:Kllpm} are symmetric under the interchange
$m \leftrightarrow -m$. Hence we deduce that $\mathbb{K}(l,l^\prime,m)
= \mathbb{K}(l,l^\prime,-m)$.
Using this we obtain
\begin{equation}
\left\langle \tilde{a}^{\,E}_{lm}\tilde{a}_{l^{\prime}m^{\prime}}^{\,E*}
\right\rangle
=  C^{EE}_{l}\delta_{ll^{\prime}}\delta_{mm^{\prime}}
+\delta_{mm^{\prime}}\, A_1
\mathbb{K}(l,l^{\prime},m)
\left(C^{EE}_{l^{\prime}}+C^{EE}_{l}\right)
.
\label{eq:EEcorrfinal}
\end{equation}
 where we have ignored the contributions proportional to 
$\delta_{l,l'}$ since they cancel out after summing over $m$.  
Similarly for the $B$ mode polarization, we obtain
\begin{equation}
\left\langle \tilde{a}^{\, B}_{lm}\tilde{a}^{\, B*}_{l^{\prime}m^{\prime}}\right\rangle
=
C_{l}^{BB}\delta_{ll^{\prime}} \delta_{mm^{\prime}} +
\delta_{mm^{\prime}}\, A_1
\mathbb{K}(l,l^{\prime},m)
\left(C^{BB}_{l^{\prime}}+C^{BB}_{l}\right)
.
\label{eq:BBcorrfinal}
\end{equation}

The dipole modulation model, Eq. \ref{eq:alpha_modu}, is very useful for 
characterizing a signal of anisotropic power that might exist in the 
polarization data. It allows an empirical parametrization of such a 
signal. Furthermore it can be used to perform simulations which are 
required for a statistical study of the anisotropy. We explicitly 
demonstrate this in the present paper. 
We also point out that an alternate model in which we may directly
introduce a dipole modulation in the $E$ mode polarization simply does
not work. 

The correlations of the $E$ and $B$ mode multipoles, Eqs.
 \ref{eq:EEcorrfinal} and
 \ref{eq:BBcorrfinal}
depend only on the parameter $A_1$ and are independent of
 $A_2$.  
Hence  
we can directly extract $A_1$ by studying the 
$E$ mode correlations.  

A similar calculation for the TE mode correlations leads to the 
following result
\begin{equation}
\left\langle \tilde  a^{\,T*}_{lm}\tilde a^E_{l'm'}\right\rangle
=C^{TE}_l \delta_{ll'}\delta_{mm'} + AC^{TE}_{l'} \xi^0_{lm;l'm'}
+{1\over 2}A_PC_l^{TE}\delta_{mm'}\mathbb{K}(l,l',m)
+{1\over 2}A^*_PC_l^{TE}\delta_{mm'}\mathbb{K}(l,l',-m)
\end{equation}
In this case also, after summing over $m$, the $\delta_{ll'}$ term
in $\mathbb{K}(l,l',m)$ and $\mathbb{K}(l,l',-m)$ drops out. Hence
it can be ignored and the remaining terms are symmetric under $m\leftrightarrow
-m$. Therefore we can express this result as
\begin{equation}
\left\langle  \tilde a^{\,T*}_{lm}\tilde a^E_{l'm'}\right\rangle
=C^{TE}_l \delta_{ll'}\delta_{mm'} + AC^{TE}_{l'} \xi^0_{lm;l'm'}
+A_1C_l^{TE}\delta_{mm'}\mathbb{K}(l,l',m)
\end{equation}

\subsection{\label{sec:evaluation}Calculation of the Polarization Correlations}

We next explicitly evaluate the integral
 in Eq. (\ref{eq:I_Integral}).  
The spin 2 harmonics \cite{Goldberg:1967} can be expressed as 
\begin{eqnarray}
{}_2Y_{lm}&=&(-1)^{m}e^{im\phi}\sqrt{\frac{(2l+1)(l-m)!(l+m)!}{4\pi(l+2)!(l-2)!}}\nonumber\\
&\times&
\sum_{r=0}^{l-2} (-1)^{l-r}
\binom{l-2}{r}\binom{l+2}{r+2-m}
\left(\sin\frac{\theta}{2}\right)^{2l-2r-2+m}\left(\cos\frac{\theta}{2}\right)^{2r+2-m}.
\label{eq:spin2Ylm}
\end{eqnarray}
The $\phi$ integration in Eq. (\ref{eq:I_Integral}) leads to  
the factor $2\pi\delta_{mm^{\prime}}.$ The $\theta$ integral is evaluated by
using the identity
\begin{equation}
\int_{0}^{\pi}d\theta
\cos\theta\sin\theta 
\sin^{m}\left(\frac{\theta}{2}\right)\cos^{n}\left(\frac{\theta}{2}\right)
=2\frac{\Gamma\left(\frac{m+2}{2}\right)\Gamma\left(\frac{n+4}{2}\right)}{\Gamma\left(\frac{m+n+6}{2}\right)}-2\frac{\Gamma\left(\frac{m+4}{2}\right)\Gamma\left(\frac{n+2}{2}\right)}{\Gamma\left(\frac{m+n+6}{2}\right)}.
\label{eq:int_irr}
\end{equation}
which can be derived by using \cite{1972hmfw.book.....A}
\[
\int_{0}^{\pi/2}\sin^{m}\theta\cos^{n}\theta
d\theta=\frac{\Gamma\left(\frac{m+1}{2}\right)\Gamma\left(\frac{n+1}{2}\right)}{2\Gamma\left(\frac{m+n+2}{2}\right)}.
\]
The function
$\mathbb{I}\left(l,m,l^{\prime},m^{\prime}\right)$
defined in Eq. \ref{eq:I_Integral} can now be expressed as
\begin{equation}
\mathbb{I}\left(l,m,l^{\prime},m^{\prime}\right)=\left(-1\right)^{l+l^{\prime}}\delta_{mm^{\prime}}{H}\left(l,l^{\prime},m\right)\mathbb{U}\left(l,l^{\prime},m\right),\label{eq:main_integral}
\end{equation}
where
\begin{equation}
\mathbb{U}\left(l,l^{\prime},m\right)=-\frac{\left(l-2\right)!\left(l+2\right)!\left(l^{\prime}+2\right)!\left(l^{\prime}-2\right)!}{2\left(l^{\prime}+l+2\right)!}\mathbb{S}\left(l,l^{\prime},m\right), \label{eq:main_function}
\end{equation}
\begin{equation}
\mathbb{S}\left(l,l^{\prime},m\right)=\sum_{r=0}^{l-2}\sum_{t=0}^{l^{\prime}-2}\mathcal{F}\left(l,l^{\prime},r,t,m\right)
\label{eq:Sllpm}
\end{equation}
and
\begin{equation}
\mathcal{F}\left(l,l^{\prime},r,t,m\right)=\frac{(-1)^{r+t}(l+l^{\prime}-r-t+m-2)!  (r+t-m+2)!   (2r+2t-2m+4-l-l^{\prime})}{r!t!   (l-2-r)!  (l^{\prime}-2-t)!
(r+2-m)!  (t+2-m)!  (l-r+m)!  (l^{\prime}-t+m)!}.
\label{eq:strange_function}
\end{equation}

%We next prove that  $\mathbb{S}\left(l,l^{\prime},m\right)$ is zero when $l^{\prime}>l+1$ or $l^{\prime}<l-1$. 

%\noindent
%Proof:

%In order to consider the case $l'<l-1$ we use the property
%\[
%\mathcal{F}\left(l,l^{\prime},r,t,m\right)=\mathcal{F}\left(l^{\prime},l,t,r,m\right)\]
%and express the function $\mathcal{F}\left(l,l^{\prime},r,t,m\right)$ as 
%\[ \frac{(-1)^{r+t}\binom{l^{\prime}+2}{l^{\prime}-t+m} \binom{l^{\prime}-2}{t}\binom{l-2}{r}} {(l-2)!(l^{\prime}+2)!(l^{\prime}-2)!}
%\left[(2r+2t-2m+4-l-l^{\prime})\prod_{s=1}^{l^{\prime}-2-t}(l-r+m+s)\prod_{v=1}^{t}(r-m+2+v)\right],\]
%In this case, we  sum over $r$ in Eq. \ref{eq:Sllpm}  keeping $t$ fixed. Using the same arguments as in the case of  $l^{\prime}>l+1$, we find that $\mathbb{S}\left(l,l^{\prime},m\right)$ is also zero for $l^{\prime}<l-1$.

 We next show that for $l'=l$,
\begin{equation}
\mathbb{S}\left(l,l^{\prime},m\right) = 
\frac{-4m\left(2l\right)!}{l\left(l+m\right)!\left(l-m\right)!\left(l+2\right)!\left(l-2\right)!}.\label{eq:result_L_L1_same}
\end{equation}

\noindent
Proof:
For $l=l^{\prime}$,  
\begin{equation}
\mathbb{S}\left(l,l^{\prime},m\right) =
\sum_{r=0}^{l-2} \frac{(-1)^{r}\binom{l+2}{l-r+m}\binom{l-2}{r}}
{(l-2)!(l+2)!(l-2)!}\, \mathbb{P}
\label{eq:sum_l_l_case}
\end{equation}
where
\begin{equation}
\mathbb{P} =
\sum_{t=0}^{l-2}(-1)^{t}\binom{l-2}{t}(2r+2t-2m+4-2l)
\times \prod_{s=1}^{l-2-r}(l-t+m+s)\prod_{v=1}^{r}(t-m+2+v)\,.
\label{eq:bbP}
\end{equation}
We next express the two products as,
$$
\prod_{s=1}^{l-2-r}(l-t+m+s)\prod_{v=1}^{r}(t-m+2+v)=(-1)^{l-2-r}\left[t^{l-2}+a_{1}t^{l-3}+a_{2}t^{l-4}+\ldots 
+a_{l-2}\right],
$$
where $a_{i}\in\mathbb{Z}$. 
Eq. \ref{eq:bbP} now becomes, 
$$
\mathbb{P} =
(-1)^{l-2-r}\sum_{t=0}^{l-2}(-1)^{t}\binom{l-2}{t}(2r+2t-2m+4-2l)
\left[t^{l-2}+a_{1}t^{l-3}+a_{2}t^{l-4}+\ldots
+a_{l-2}\right].\label{eq:sum_t_L_L_case}
$$
By using Eq. \ref{eq:fnpfinal} we find that  
only two terms, i.e. those proportional to $t^{l-1}$ and $t^{l-2}$, 
contribute. Thus we obtain 
\begin{equation}
\mathbb{P} =
(-1)^{l-2-r}\sum_{t=0}^{l-2}(-1)^{t}\binom{l-2}{t}
\left[(2r-2m+4-2l+2a_{1})t^{l-2}+2t^{l-1}\right].\label{eq:t_sum_l_l}
\end{equation}
The constant $a_1$ can be determined by using the result that if
$$
\prod_{i=1}^{n}(x+\alpha_{i})=x^{n}+x^{n-1}a_{1}+\ldots a_{n}\,,
$$
then $a_{1}=\sum_{i=1}^{n}\alpha_{i}.$ Thus we obtain
$$
a_{1}=\sum_{s=1}^{l-2-r}(-l-m-s)+\sum_{v=1}^{r}(-m+2+v)=
{1\over 2}
\left[-3l^{2}+7l-2ml+4m-2+2r(2l+1)\right].
$$
Using Eq. \ref{eq:fnpfinal} we can express 
Eq. \ref{eq:t_sum_l_l} as, 
$$
\mathbb{P} =
(-1)^{-r}(l-2)!\left[-2l^{2}+2l-2ml+2m+4r(l+1)\right].
$$
Substituting in Eq. \ref{eq:sum_l_l_case} we obtain
\[
\mathbb{S}\left(l,l^{\prime},m\right) =
\sum_{r=0}^{l-2}\binom{l+2}{l-r+m}\binom{l-2}{r}\frac{\left[-2l^{2}+2l-2ml+2m+4r\left(l+1\right)\right]}{\left(l+2\right)!\left(l-2\right)!}.
\]
This sum can be divided into two parts
\[
\frac{\left(-2l^{2}+2l-2ml+2m\right)}{\left(l+2\right)!\left(l-2\right)!}\sum_{r=0}^{l-2}\binom{l+2}{l-r+m}\binom{l-2}{r}+\frac{4\left(l+1\right)}{\left(l+2\right)!\left(l-2\right)!}\sum_{r=0}^{l-2}r\binom{l+2}{l-r+m}\binom{l-2}{r},
\]
In the second sum $r=0$ does not contribute. Hence after some simplifications,
 it can be re-expressed as,
\[
\frac{4\left(l-2\right)(l+1)}{\left(l+2\right)!\left(l-2\right)!}\sum_{t=0}^{l-3}\binom{l+2}{l-t-1+m}\binom{l-3}{t}.
\]
We can evaluate both of these sums by using the Vandermonde Convolution
property of binomial coefficients \cite{Gallier:2011} which can be stated as, 
\[
\sum_{k=0}^{m}\binom{m}{k}\binom{p}{n-k}=\binom{m+p}{n},\ m+p\ge n\ \&\ m,\ n,\ p\ge0\,.
\]
We finally obtain 
\begin{eqnarray}
\mathbb{S}\left(l,l^{\prime},m\right) &=&
\frac{\left(-2l^{2}+2l-2ml+2m+4\right)\left(2l\right)!}{\left(l+m\right)!\left(l-m\right)!\left(l+2\right)!\left(l-2\right)!}+\frac{4\left(l+1\right)\left(l-2\right)\left(2l-1\right)!}{\left(l+2\right)!\left(l-2\right)!\left(l-m\right)!\left(l-1+m\right)!}\nonumber\\
&=&\frac{-4m\left(2l\right)!}{l\left(l+m\right)!\left(l-m\right)!\left(l+2\right)!\left(l-2\right)!}\,,
\end{eqnarray}
which is the desired result and leads to 
 $\mathbb{U}(l,l',m)$ given in Eq. \ref{eq:Kllpm} for the case $l=l'$.

We next show that
\begin{equation}
\mathbb{S}\left(l,l^{\prime},m\right)=-
\begin{cases}
\frac{2(2l)!}{(l+m)!(l-m)!(l+2)!(l-2)!} &
l^{\prime}=l+1\\
\frac{2(2l^{\prime})!}{(l^{\prime}+m)!(l^{\prime}-m)!(l^{\prime}+2)!(l^{\prime}-2)!}
& l^{\prime}=l-1
\end{cases}
\end{equation}

\noindent
Proof:
We first consider the case $l^{\prime}=l+1$.  
We can write Eq. (\ref{eq:strange_function}) as
\begin{eqnarray}
\mathcal{F}\left(l,l^{\prime},r,t,m\right)&=&(-1)^{r+t}
\binom{l+2}{l-r+m}
\binom{l^{\prime}-2}{t} 
\frac{\left(2r+2t-2m+4-l-l^{\prime}\right)}{\left(l^{\prime}-2\right)!\left(l+2\right)!} \nonumber\\ 
&\times &\left[\frac{(l+l^{\prime}-r-t+m-2)!}{(l-r-2)!(l^{\prime}-t+m)!}\right]
\left[\frac{(r+t-m+2)!}{r!(t+2-m)!}\right].
\label{eq:Fllprtm}
\end{eqnarray}
After simplification of the terms in the two square brackets, this becomes
\[\frac{(-1)^{r+t}\binom{l+2}{l-r+m}\binom{l-2}{r}\binom{l^{\prime}-2}{t}} {(l^{\prime}-2)!(l+2)!(l-2)!}
\left[(2r+2t-2m+4-l-l^{\prime})\prod_{s=1}^{l-2-r} (l^{\prime}-t+m+s)\prod_{v=1}^{r}(t-m+2+v)\right]. \]
We can write the term in the square brackets above as $a_{0}t^{l-1}+a_{1}t^{l-2}\ldots a_{l-1}$, where $a_{i}\in\mathbb{Z}$.  Keeping $r$ fixed, the sum over $t$ in Eq. \ref{eq:Sllpm} yields,
\begin{equation}
\sum_{t=0}^{l^{\prime}-2}\mathcal{F}\left(l,l^{\prime},r,t,m\right)
= \sum_{t=0}^{l^{\prime}-2}(-1)^{t}\binom{l^{\prime}-2}{t}
\left[a_{0}t^{l-1}+a_{1}t^{l-2}+ ... +a_{l-1}\right].\label{eq:t_sum_first}
\end{equation}
By using Eq. \ref{eq:Fllprtm} on the left hand side 
and by comparing both sides, we find that $a_{0}$
is equal to $2$$\left(-1\right)^{l-2-r}$.
Using the second case of Eq. \ref{eq:fnpfinal} we obtain 
$$
\sum_{t=0}^{l^{\prime}-2}\mathcal{F}\left(l,l^{\prime},r,t,m\right)
= -2\left(-1\right)^{-r}\left(l-1\right)!=-2\left(-1\right)^{r}\left(l-1\right)!
$$
Finally
the sum over $r$, after simplification, yields 
$$
\mathbb{S}\left(l,l^{\prime},m\right)=-
\frac{-2}{(l+2)!(l-2)!}\left[\sum_{r=0}^{l-2}\binom{l+2}{l-r+m}\binom{l-2}{r}\right]=\frac{-2(2l)!}{(l+2)!(l-2)!(l+m)!(l-m)!},
$$
where we have again used the Vandermonde Convolution property. 
A similar analysis for
the case $l^{\prime}=l-1$
yields
$$
\mathbb{S}\left(l,l^{\prime},m\right)=-
\frac{-2 (2l^{\prime})!} {(l^{\prime}+2)! (l^{\prime}-2)!
(l^{\prime}+m)! (l^{\prime}-m)!}.
$$
These lead to the result for  
 $\mathbb{U}(l,l',m)$ given in Eq. \ref{eq:Kllpm}
for the cases
$l'=l\pm 1$

\section{Data analysis}
For the case of temperature data, we perform a detailed analysis of the
signal. This allows us to obtain 
updated results for the statistic $S_H^{TT}$ with the 2015 Planck data. 
We studied this statistic  
in the multipole ranges
 $2\le l \le 64$, $30\le l \le 64$ and $30\le l \le 100$ 
 in the Planck 
2015 CMB intensity maps. The dipole modulation signal in the CMB 
temperature map was observed in the multipole range $2-64$ \cite{Rath:2014}.

For the case of CMB polarization a detailed analysis is not possible
at this stage since the data for low $l$ is unreliable. Furthermore
we are unable to properly simulate the noise corresponding to PLANCK 
detectors. Hence, even for large $l$, it is not possible for us to 
obtain a reliable estimate of the errors and the significance of
the signal. For this reason we confine ourselves to a preliminary 
analysis of the polarization signal. However it
may still serve a useful purpose in revealing the preferred direction
indicated by data. 
We searched for modulation signal in the multipole ranges $40-100$,
$40-125$, 
$50-100$, $50-125$, $50-150$ and $50-200$.   
The lower limit of $l=40$ was chosen since the data for lower $l$ is so
far poorly understood.

\subsection{Planck 2015 temperature data analysis}
We have performed our analysis on both Commander and SMICA IQU maps. For 
SMICA maps, we have performed the analysis after masking the maps and then inpainting the masked maps using the MRS package of iSAP software or alternatively masking and then filling the masked portion of the map with isotropic data generated using the CAMB simulation package with Planck 2015 parameter set. The data analysis was performed with HEALPix software \cite{Gorski:2004}.

\begin{figure}[t]
    \centering
    \includegraphics[width=1.0\textwidth]{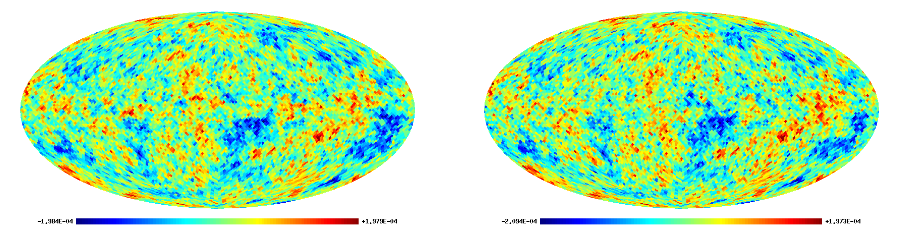}
    \caption{Left: SMICA inpainted temperature map. Right: SMICA 
temperature map with the masked portions filled by isotropic
randomly generated data. Both maps are for $N_{\text{side}} = 32$ 
and the temperature is given in units of K.}
    \label{Figure_maps}
\end{figure}

The analysis of the Commander and SMICA inpainted maps was performed identically. 
The maps were downgraded, without any smoothing to either $N_{\text{side}}=32$, for multipole ranges $2-64$ and $30-64$, or $N_{\text{side}}=64$, for $30-100$
multipole range and the analysis was performed on these downgraded maps. 
We removed the dipole and monopole from CMB intensity maps and calculated the quantities $\sum_lC^{TT}_{l,l+1}$ and $\sum_ll(l+1)C^{TT}_l$ over the aforementioned multipole ranges. 

We also used another method to analyze the 
SMICA maps. In this case the SMICA I maps were first masked with their 
respective masks. The pixels which were masked were filled with data from I 
simulated maps generated using lensed scalar $C_l$ values generated with the  CAMB Boltzmann solver \cite{Lewis:1999} for the Planck 2015 parameters \cite{Ade:2015}, and the Synfast program from the HEALPix package. The masked and filled I maps 
have $N_{\text{side}} = 2048$, same as that of the original maps. 
These maps were smoothed with a FWHM equal to 3 times the pixel size of
the low resolution map with $N_{\text{side}} = 256$  before downgrading
the map to remove the discontinuities at the mask boundary. The smoothed maps are then degraded to $N_{\text{side}} = 256$ followed by a further degradation to $N_{\text{side}} = 32$ or $N_{\text{side}}=64$, depending on the multipole range, for final analysis. We generated 100 such filled maps and 
the individual results were found to depend on the random realization 
used to fill 
the masked portions of the sky. The results presented here are the mean values  for the statistic $S^{TT}_H$ and direction for which it maximizes.
The SMICA I maps masked and reconstructed using these two procedures,
i.e. inpainting and random filling, are shown in Fig. \ref{Figure_maps}.

\begin{table}
	\begin{center}
		\begin{ruledtabular}
	    \begin{tabular}{ p{2.5 cm} c c c c }
	    Map & $S^{TT}_H$ in $10^{-2}$ $\text{mK}^2$ & $A$ & $(l, b)$ & P-value \\  \hline
	    Commander & $2.55\pm0.68$ & $0.082 \pm 0.018 $ & $(232^\circ\pm18^\circ, -14^\circ\pm18^\circ)$ & $0.20 \%$ \\ %\hline
	    SMICA(inp.) & $2.39\pm0.70$ & $0.069 \pm 0.013$ & $(236^\circ\pm27^\circ, -11^\circ\pm20^\circ)$ & $0.70 \%$ \\ %\hline
	    SMICA(filled) & $2.44\pm0.71$ & $0.078\pm0.019$ &   $(242^\circ\pm16^\circ, -17^\circ\pm20^\circ)$ &$0.50 \% $  \\
%	    Bias corrected & $2.36\pm0.71$ & $0.078\pm0.019$ &   $(242^\circ\pm16^\circ, -17^\circ\pm20^\circ)$ & $0.85 \%$  \\
	    \end{tabular}
	    \caption{The maximum TT Mode $S_H$  values along with the dipole modulation parameter $A$,  the preferred direction of maximization and the P-value.% Bias corrected value for filled SMICA maps provided in the last row.
}
	    \label{Table_TT}
	    \end{ruledtabular}
	\end{center}
\end{table}

We have fitted the TT mode values of $S_H$ in order to extract the
 value of the dipole modulation amplitude $A$ of Eq. (\ref{modelz}). We simulated 100 isotropic CMB maps using Planck 2015 parameters with $N_{\text{side}} = 512$ ($N_{\text{side}} = 1024$ for SMICA filled analysis). These maps were rotated to have the z axis pointing along the direction of maximum statistic and they were modulated using 
Eq. (\ref{eq:model}). The modulated maps were downgraded to $N_{\text{side}} = 32$. Each of the downgraded simulated maps were fitted for the value of $A$ that would give the value of $S_H$ closest to the one observed in the data along the direction along which it maximizes in the Planck 2015 CMB maps. 
We averaged over 100 best fit values of $A$ obtained by this method giving the modulation amplitude for the results given in Table \ref{Table_TT}.

To estimate the error in $S_H^{TT}$ we generated 1000 maps at $N_{\text{side}}=512$ ($N_{\text{side}} = 1024$ for SMICA filled analysis) and modulated 
the maps with the best-fit value of $A$ along the direction of maximum statistic using relation (\ref{modelz}). The modulated CMB maps were downgraded to $N_{\text{side}}= 32$ or $N_{\text{side}}=64$ depending on the multipole range
under consideration and $S_H^{TT}$ was calculated along the direction of modulation. The standard deviation of the 1000 simulated maps gives the error in $S_H^{TT}$. For the masked and filled SMICA maps, the process of filling the masked region of the SMICA maps with isotopic data is expected to introduce bias in the obtained results for $S_H$. This bias correction is relatively small \cite{Rath:2014} and
we ignore it in our analysis. It is expected to enhance the signal by about 
8\%.

The error estimation in the preferred direction was performed by simulating 50 isotropic temperature maps. We modulated these maps with the best-fit value of $A$ along the observed direction in the data. The direction along which the statistic $S_H^{TT}$ maximises in the maps was found and the standard deviation of the coordinates gave corresponding errors.

\begin{figure}[t]
    \centering
    \includegraphics[width=1.0\textwidth]{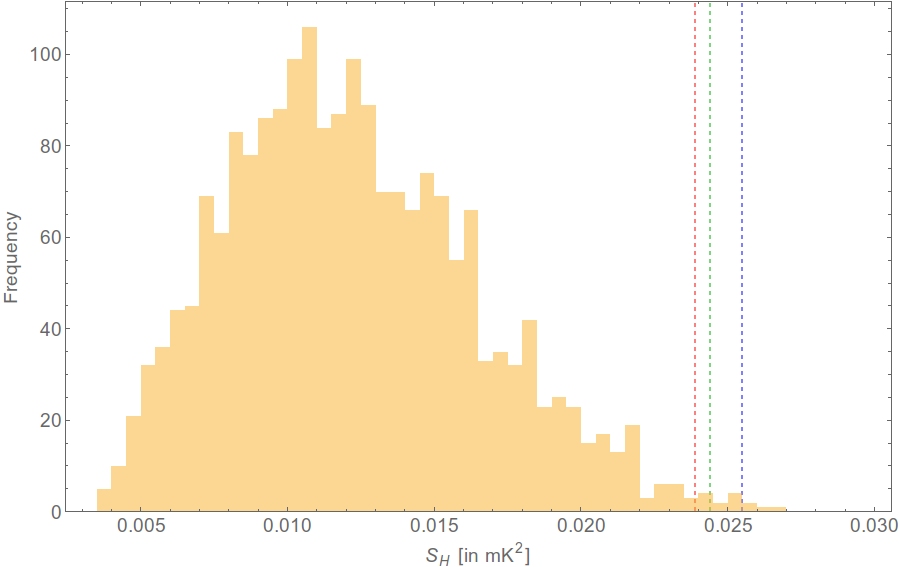}
    \caption{Histogram of TT $S_H$ values obtained by using isotropic $\Lambda$CDM simulations. The observed $S_H$ values are indicated by dashed lines. 
The results obtained by using the Commander map (blue line), SMICA inpainted
(red line) and  
SMICA filled map (green line) are shown. }
    \label{Figure_distribution}
\end{figure}

Finally to test the significance of our results we simulate 2000 isotropic CMB maps for $2-64$ multipole range and 500 maps for the other multipole ranges, with Planck 2015 parameters and search for the direction along which the statistic maximizes. The values of $S_H$ obtained by this process is used to obtain probability distribution of the statistic $S_H$ for the isotropic hypothesis. 
The histogram is given in Fig. \ref{Figure_distribution}. P-values for individual results were obtained as the percentage of simulation results that equal or exceed the observed result for the statistic. 

\begin{table}
	\begin{center}
	\begin{ruledtabular}
	    \begin{tabular}{c c c c c c }
	    $l$ Range & $S^{TT}_H$ in $10^{-2}$ $\text{mK}^2$ &$A$& $(l, b)$ & P-value & $R$\\ \hline
	    2-64 & $2.55 \pm 0.68$ & $0.082\pm0.018$& $(232^\circ\pm 18^\circ, -14^\circ\pm 18^\circ)$ &  $0.20 \%$  & $0.065$\\ %\hline
	    30-64 & $1.00\pm 0.43$ & $0.052\pm0.019$& $(194^\circ\pm25^\circ, -4^\circ\pm24^\circ)$ & $66.6 \%$ & $0.040$ \\  %\hline
	    30-100  &$0.91\pm0.72$ & $0.018\pm0.011$& $(277^\circ\pm81^\circ, 4^\circ\pm30^\circ)$ & $83.0 \%$ & $0.013$ \\
	    \end{tabular}
	    \caption{TT mode $S_H$ results for Planck Commander 2015 maps in different multipole ranges. }
	    \label{Table_TT3}
	\end{ruledtabular}    
	\end{center}
\end{table}

\subsection{Planck 2015 polarization data analysis}
For the case of polarization, as explained above, we are unable to 
perform a detailed analysis of dipole modulation. Here we confine ourselves
to simply making an estimate of the statistic $S_H^{EE}$ and the 
corresponding preferred direction using the Commander map. Since the low
$l$ multipoles are not expected to be reliable we confine our study
to the multipole ranges $40-100$, $40-125$, $50-100$ and $50-125$.
 Furthermore we use the dipole
modulation model for polarization 
in order to generate simulated maps which display polarization
power anisotropy and to determine the distributions of the
corresponding statistic $S_H$ for the $E$ mode polarization.
 The main purpose of this study is to illustrate 
the utility of this model.

\section{Results}
In this section we first present the results for the temperature analysis 
and later those of polarization.

\subsection{Temperature}
The TT mode results for multipole range $2\le l \le 64$ are summarized 
in Table \ref{Table_TT}. As stated in the previous section the dipole modulation in the CMB temperature signal is present in lower multipoles up to $l = 64$. Our primary TT mode results are for this multipole range. We can compare our results with \cite{Rath:2014} to look for changes between the Planck 2013 and Planck 2015 data. The results for SMICA filled maps for Planck 2013 data are: $S_H^{TT}=(2.1\pm0.5)\times10^{-2}\text{mK}^2$, a bias corrected value of $(2.3\pm0.6)\times10^{-2}\text{mK}^2$, along $(229^\circ,-16^\circ)$ with $A=0.074\pm0.019$. Comparing the 2013 and 2015 SMICA filled map results we notice agreement in both the values of $S_H^{TT}$ and direction of maximization, while the value of $A$ is also comparable within the error limits. The results for SMICA inpainted map for Planck 2013 data are: $S_H^{TT}=(2.7\pm0.7)\times10^{-2}\text{mK}^2$, along $(232^\circ,-12^\circ)$. For SMICA inpainted maps too, we find good agreement with previous results. We also note that the 2015 results have smaller P-values when compared with isotropic $\Lambda$CDM simulations generated using 2015 Planck parameters. We can compare our results with Planck Collaboration's analysis of dipole modulation \cite{Ade:2015XVI}.
Planck 2015 best fit values of modulation amplitude $A$ is $0.063^{+0.025}_{-0.013}$ for Commander maps and $0.062^{+0.026}_{-0.013}$ for SMICA map. The direction of modulation is $(213^\circ,-26^\circ)\pm 28^\circ$ for both SMICA and Commander maps. Both the modulation amplitude and the direction agree with our results within the quoted errors. 
 Our results are also consistent with those  obtained 
in \cite{Aiola:2015}. 

\begin{figure}[t]
    \centering
    \includegraphics[width=1.0\textwidth]{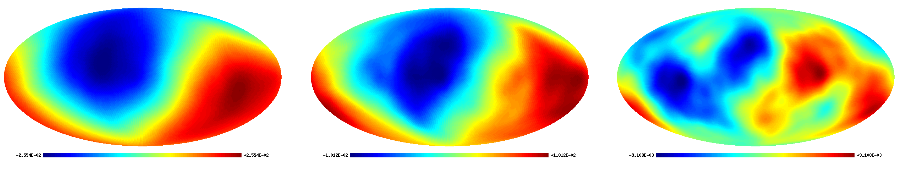}
    \caption{From the left $S_H^{TT}$ sky maps for Commander data in multipole ranges $2-64$, $30-64$ and $30-100$. }
    \label{Figure_modulation}
\end{figure}

It has been found by an earlier  analysis that the hemispherical anisotropy
effect rapidly dies out when the lower multipoles are excluded or when summed to higher multipoles \cite{Donoghue:2005,Hanson:2009,Quartin:2014,Ade:2015XVI}. 
In order to study the multipole dependence we also investigate the effect
in the multipole ranges $30-64$ and $30-100$. 
The $S_H^{TT}$ obtained in 30-64 for Commander map is contained in Table \ref{Table_TT3}. In this case we do not find a significant signal of
$l, l+1$ correlation.  The 
temperature map $S_H$ is very much  in agreement with expectations from 
isotropic theory. We also notice that 
 the $S_H^{TT}$ maximizes along directions which change from one range to another. This can also be seen from Fig. \ref{Figure_modulation}. These maps 
show $S_H^{TT}$ for different directions in ranges 2-64, 30-64 and 30-100. 
It can be seen that left to right the dipole pattern becomes less distinct.

The SMICA inpainted TT mode results are as follows:
\begin{itemize}
\item In the range 30-64, $S_H^{TT}$ maximizes at $(1.24\pm0.43)\times 10^{-2}\text{mK}^2$ along $(191^\circ\pm 27^\circ,-4^\circ\pm 28^\circ)$ with a P-value of $45 \%$ and $A=0.059\pm0.019$.
\item For 30-100 range, $S_H^{TT}$ maximizes at $(1.45\pm0.71)\times 10^{-2}\text{mK}^2$ along $(207^\circ\pm 38^\circ,-18^\circ\pm 27^\circ)$ with a P-value of $41.2 \%$ and $A=0.027\pm0.013$.
\end{itemize}

The SMICA filled map temperature results are:
\begin{itemize}
\item In 30-64 range, $S_H^{TT}$ maximizes at $(1.00\pm0.43)\times 10^{-2}\text{mK}^2$ along $(228^\circ\pm 28^\circ,-9^\circ\pm 27^\circ)$ with a P-value of $66 \%$ and $A=0.055\pm0.021$.
\item  The 30-100 range $S_H^{TT}$ maximizes at $(1.64\pm0.77)\times 10^{-2}\text{mK}^2$ along $(238^\circ\pm 34^\circ,-11^\circ\pm 28^\circ)$ with a P-value of $27 \%$ and $A=0.031\pm0.014$.
\end{itemize} 
 Both SMICA inpainted and filled maps results show similar patterns as Commander map results.

\subsection{Polarization}
In this section we present our results for polarization. 
As explained
above, we do not make an attempt to compute either the errors or the 
significance of this effect due to uncertainties in the noise simulation
in polarization. However we do present the results of a simulation
in order to illustrate the utility of the polarization dipole
modulation model, Eq. \ref{eq:alpha_modu}. 

In Table \ref{Table_EE} we give the results for $S_H^{EE}$, the preferred
direction and the ratio $R$ for the $E$ mode polarization. 
The result for the case of the multipole range $40-100$ is mildly
interesting. This is because the preferred direction aligns closely 
with the CMBR dipole. 
A preferred direction similar to the one obtained in the range 
$40-100$ has been seen
in many other studies \cite{Ralston:2004,Schwarz:2004}, including the 
radio polarization dipole axis \cite{Jain:1998}, the CMB quadrupole -- octopole alignment axis \cite{deOliveiraCosta:2003}, the NVSS dipole \cite{Singal:2011,Gibelyou:2012,Rubart:2013,Kothari:2013}, 
the radio polarization flux dipole axis \cite{Tiwari2015}
as well as the optical polarizations from distant quasars 
\cite{Hutsemekers:1998,Hutsemekers:2000fv,Jain:2003sg,Ralston:2004}. 
It is possible that the present signal in $E$ mode polarization in 
this range arises from some residual contamination of the low $l$ noise
systematic bias. Alternatively it may be a signal of some astrophysical
or cosmological effect. This may be settled by future refinements in data.

\begin{figure}[t]
    \centering
    \includegraphics[width=1.0\textwidth]{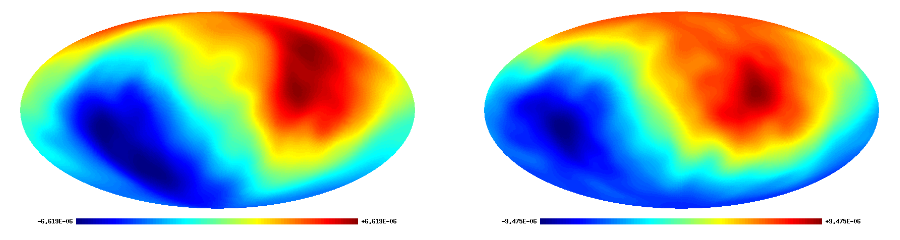}
    \caption{The $S_H^{EE}$ sky maps for Commander data in multipole ranges 
$40-100$ (left) and $40-125$ (right). }
    \label{Figure_modulationpol}
\end{figure}

For the higher multipole ranges the preferred direction starts to deviate. 
For example, in the multipole range $40-125$, $50-100$, $50-125$ it lies
closer to the galactic plane. 
This might be an indication that it is moving closer to the axis obtained
in the case of temperature. Alternatively, since it lies very close to
the galactic plane, the signal in this range might be dominated by
foregrounds. 
The plots of $S_H^{EE}$ for the multipole range $40-100$ and $40-125$ are shown
in Fig. \ref{Figure_modulationpol}.

\begin{table}
	\begin{center}
	\begin{ruledtabular}
	    \begin{tabular}{ c c c c }
	    $l$ Range & $S^{EE}_H$ in $10^{-6}$ $\text{mK}^2$ & $(l, b)$ & $R$\\ \hline
	    40-100 & 6.6 &  $(260^\circ, 44^\circ)$ &    $0.031$\\ %\hline
	    40-125 & 9.4 &  $(291^\circ, 14^\circ)$ &    $0.023$\\ %\hline
	    50-100 & 6.6 &  $(286^\circ, 15^\circ)$ &  $0.033$ \\  %\hline
	    50-125 & 9.5 &  $(291^\circ, 14^\circ)$ &  $0.025$ \\  %\hline
	    \end{tabular}
	    \caption{EE mode $S_H$ results for Planck Commander 2015 maps in different multipole ranges. 
}
	    \label{Table_EE}
	\end{ruledtabular}
	\end{center}
\end{table}

%\begin{figure}[t]
%    \centering
%    \includegraphics[width=1.0\textwidth]{modulation_updated.png}
%    \caption{Clockwise from top left: The $S_H^{EE}$ sky maps for 
%Commander data in multipole ranges 
%$50-100$, $50-125$, $50-150$ and $50-200$.  }
%    \label{Figure_modulationpol50}
%\end{figure}

\begin{figure}[t]
    \centering
    \includegraphics[width=1.0\textwidth]{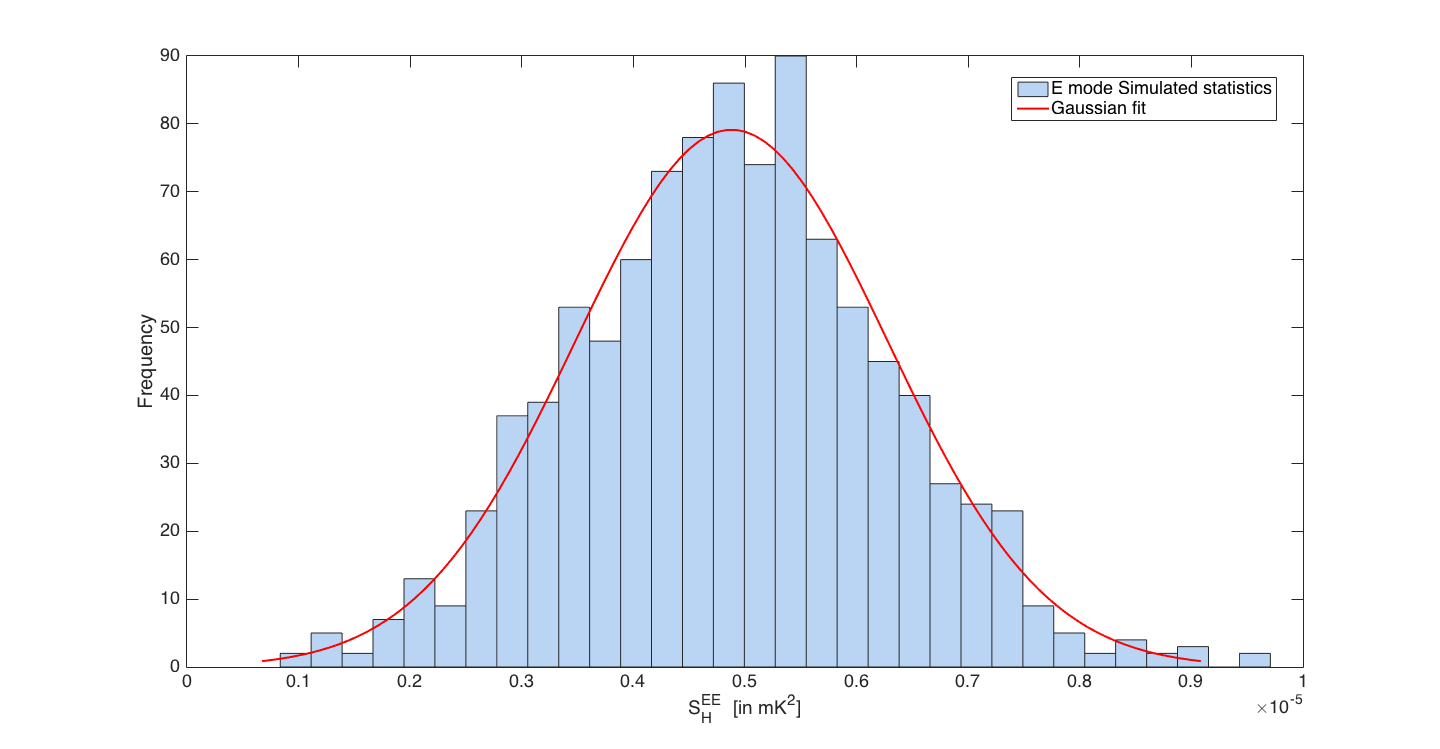}
    \caption{The $S_H^{EE}$ histogram for the simulated 
dipole modulated $E$ mode polarization maps with the parameter $A_1=0.05$. }
    \label{Figure_SH_hist1}
\end{figure}

We next generate 1000 simulated $Q$ and $U$ mode polarization maps which
include dipole modulation with parameter $A_1=0.05$ for
the multipole range $40-100$. We 
compute the statistics $S_H^{EE}$ for these simulated
maps. The resulting distribution of this statistics is shown in 
Fig. \ref{Figure_SH_hist1}.
We find that
 the distribution is close to normal. 
The parameter chosen is not too far from what might be required 
to fit the $E$ mode results given in Table \ref{Table_EE}.

\section{Conclusions}
This work has updated the previous results for dipole modulation in CMB temperature field. 
We find that the dipole modulation of the temperature field, observed in 
WMAP data and Planck 2013 data is also present in Planck 2015 data. 
We have computed the quantity $S_H^{TT}$ and the results are found to be
in agreement with the values obtained from the previous data \cite{Rath:2014}. 
Our result for modulation parameter $A$ and modulation direction are compatible with Planck 2015 results \cite{Ade:2015XVI}.

We have presented a preliminary analysis of the dipole modulation in
the CMBR polarization. In this case we confine our analysis only to the
multipole values $l>40$ since the lower $l$ values are not expected to
be reliable. The detector noise is expected to make considerable 
contribution in this case. Since this information is not available
to us, we do not attempt to compute the significance of
the signal or the associated error in extracted parameters.
In any case our preliminary 
investigation reveals some interesting results about the direction
of the signal. We find that for the range of multipoles $40-100$
the preferred direction is close to the CMBR dipole and for a slightly
higher range $50-100$ or $40-125$, the direction shifts closer to
the galactic plane. 
The direction found for the lower range of multipoles is somewhat interesting
since a similar direction has been found in several other data sets
\cite{Ralston:2004,Schwarz:2004,Jain:1998,deOliveiraCosta:2003,Singal:2011,Gibelyou:2012,Rubart:2013,Kothari:2013,Tiwari2015}.

For the case of polarization we also propose
a dipole modulation model, Eq. \ref{eq:alpha_modu}. 
This is a useful model
which generalizes the temperature dipole modulation \cite{Gordon2005,Gordon2007,Prunet2005,Bennett2011},  Eq. \ref{eq:model}, to the case of polarization.
We find that this model leads to correlations between $l$ and $l+1$ multipoles
of the polarization field. We determine the form of these correlations for
the case of $EE$, $BB$ and $TE$ fields.
We show the utility of this model by creating simulated polarization
maps with non-zero values of the dipole modulation parameter, 
$A_1$.  We find that distribution of the resulting
statistic for the case of $E$ model polarization is well described by
a Gaussian. 
We expect that this dipole modulation model
and our proposed analysis procedure may be very useful in the 
study of anisotropy which might exist in the CMBR polarized field. 

\section{\label{sec:Appendix}Appendix}

In this Appendix we
 derive the mathematical results required in proving the results
in section \ref{sec:evaluation}. 
Let $f\left(n,p\right)$ be defined as 
\begin{equation}
f\left(n,p\right)=\sum_{r=0}^{n}r^{p}\left(-1\right)^{r}\binom{n}{r}
\label{eq:func_defi}
\end{equation}
where $n,\ p\ge0$. This function satisfies the recurrence
relation
\begin{equation}
f(n,p+1)=\begin{cases}
(-n)\sum_{q=0}^{p}f\left(n-1,q\right)\binom{p}{q} & n\ne0\\
0 & n=0
\end{cases}.
\label{eq:rec_relation}
\end{equation}

\noindent
Proof:
The result for $n=0$ can be verified by direct substitution.
For $n>0$, we can write $f(n,p+1)$ as 
$$
f(n,p+1)=
n\sum_{r=0}^{n}(-1)^{r}r^{p}\frac{(n-1)!}{(n-r)!(r-1)!}.
$$
On the right hand side the term corresponding to $r=0$ is zero  
 due to the factor $\left(r-1\right)!$ 
in the denominator. Hence we can
 start the sum from $r=1$. Setting $r-1=t$, 
we obtain
$$
f(n,p+1)=-n\sum_{t=0}^{n-1}(-1)^{t}(1+t)^{p}\frac{(n-1)!}{\left(n-t-1\right)!t!}=-n\sum_{q=0}^{p}\binom{p}{q}\sum_{t=0}^{n-1}\left(-1\right)^{t}\binom{n-1}{t}t^{q},
$$
where we have used the binomial theorem.
Furthermore
$$
f(n-1,q)=\sum_{t=0}^{n-1}(-1)^{t}\binom{n-1}{t}t^{q},
$$
and hence
$$
f(n,p+1)=-n\, \sum_{q=0}^{p}f(n-1,q)\binom{p}{q}.
$$
This proves the result in 
Eq. \ref{eq:rec_relation} for $n\ne 0$. 

The function $f\left(n,p\right)$ 
defined in Eq.  
\ref{eq:func_defi} is given by
\begin{equation}
f(n,p)=\begin{cases}
0 & p<n,\ n\ne0\\
(-1)^{n}n! & p=n,\ n\ge0\\
\frac{n(n+1)}{2}(-1)^{n}n! & p=n+1,\ n\ge0
\end{cases}
\label{eq:fnpfinal}
\end{equation}

\noindent
Proof:
The first part can be proven by using the Corollary 2 of \cite{Ruiz:1996} with
$x=0$ and $p=n-j$. The result follows since  $p<n$ when 
 $1\le j\le n$. 
Furthermore the second part can be obtained by using 
 Theorem 1 of \cite{Ruiz:1996} 
with $x=0$.

Finally we consider the last part. By using Eq. \ref{eq:rec_relation}
we find that
for $n=0$, $p=n+1$, 
$f\left(0,p\right)=0$, which agrees with the result given in Eq. 
\ref{eq:fnpfinal}.  
 We next consider $n>0$. By using Eq.  
\ref{eq:rec_relation} we obtain 
$$
f(n,n+1)=-n\sum_{q=0}^{n}f(n-1,q)\binom{n}{q}
=-n\left[f(n-1,n-1)\binom{n}{n-1}+f(n-1,n)\right].
$$
That is, the only nonvanishing terms in this sum are obtained by setting 
 $q=n-1$ and $q=n$. 
Using the second case of Eq. 
\ref{eq:fnpfinal} this can be
further simplified in the form of the following
recurrence relation:
\begin{equation}
P(n)=-n\left[-(-1)^{n}n!+P(n-1)\right]\,,
\label{eq:induction_recurrence}
\end{equation}
where $P\left(n\right)=f\left(n,n+1\right)$. 
We next 
 proceed by induction. For $n=1$, by using \ref{eq:func_defi}
we obtain
$$
P(1)=f(1,2)=-1,
$$
which agrees with the result given in Eq. 
\ref{eq:fnpfinal}.
 We next assume that this result is true for $n=k$, i.e.,
$$
P(k)=\frac{k(k+1)}{2}(-1)^{k}k!
$$
and show that 
it is also true for $P\left(k+1\right)$.
By using recurrence relation \ref{eq:induction_recurrence} we  
obtain
$$
P(k+1)=-(k+1)\left[-(-1)^{k+1}(k+1)!+\frac{k(k+1)}{2}(-1)^{k}k!\right]
=\frac{(k+1)(k+2)}{2}(-1)^{k+1}(k+1)!
$$
which agrees with Eq. \ref{eq:fnpfinal}.
Hence the third part is also proven by induction.

\section*{Acknowledegments}
Rahul Kothari sincerely acknowledges CSIR, New
Delhi for the award of fellowship during the work.
Some of the results in this paper have been derived using HealPix package
\citep{Gorski:2005}. We used standard Boltzmann solver CAMB (http://camb.info/readme.html) for our
theoretical calculations and acknowledge the use of PLANCK data
available from NASA LAMBDA site (http://lambda.gsfc.nasa.gov).
We are very grateful to Anthony Banday and Yashar Akrami for
explaining the limitations of the PLANCK polarization data. Finally
we thank J. L. Starck for a useful correspondence.

\bibliography{DipoleModulation}

%Merlin.mbs v4.21 2009-07-09.
\begin{thebibliography}{10}%
\makeatletter
\providecommand \@ifxundefined [1]{%
 \ifx #1\undefined \expandafter \@firstoftwo
 \else \expandafter \@secondoftwo
\fi
}%
\providecommand \@ifnum [1]{%
 \ifnum #1\expandafter \@firstoftwo
 \else \expandafter \@secondoftwo
\fi
}%
\providecommand \enquote [1]{``#1''}%
\providecommand \bibnamefont  [1]{#1}%
\providecommand \bibfnamefont [1]{#1}%
\providecommand \citenamefont [1]{#1}%
\providecommand\href[0]{\@sanitize\@href}%
\providecommand\@href[1]{\endgroup\@@startlink{#1}\endgroup\@@href}%
\providecommand\@@href[1]{#1\@@endlink}%
\providecommand \@sanitize [0]{\begingroup\catcode`\&12\catcode`\#12\relax}%
\@ifxundefined \pdfoutput {\@firstoftwo}{%
 \@ifnum{\z@=\pdfoutput}{\@firstoftwo}{\@secondoftwo}%
}{%
 \providecommand\@@startlink[1]{\leavevmode\special{html:<a href="#1">}}%
 \providecommand\@@endlink[0]{\special{html:</a>}}%
}{%
 \providecommand\@@startlink[1]{%
  \leavevmode
  \pdfstartlink
   attr{/Border[0 0 1 ]/H/I/C[0 1 1]}%
   user{/Subtype/Link/A<</Type/Action/S/URI/URI(#1)>>}%
  \relax
 }%
 \providecommand\@@endlink[0]{\pdfendlink}%
}%
\providecommand \url  [0]{\begingroup\@sanitize \@url }%
\providecommand \@url [1]{\endgroup\@href {#1}{\urlprefix}}%
\providecommand \urlprefix [0]{URL }%
\providecommand \Eprint[0]{\href }%
\@ifxundefined \urlstyle {%
  \providecommand \doi [1]{doi:\discretionary{}{}{}#1}%
}{%
  \providecommand \doi [0]{doi:\discretionary{}{}{}\begingroup
  \urlstyle{rm}\Url }%
}%
\providecommand \doibase [0]{http://dx.doi.org/}%
\providecommand \Doi[1]{\href{\doibase#1}}%
\providecommand \bibAnnote [3]{%
  \BibitemShut{#1}%
  \begin{quotation}\noindent
    \textsc{Key:}\ #2\\\textsc{Annotation:}\ #3%
  \end{quotation}%
}%
\providecommand \bibAnnoteFile [2]{%
  \IfFileExists{#2}{\bibAnnote {#1} {#2} {\input{#2}}}{}%
}%
\providecommand \typeout [0]{\immediate \write \m@ne }%
\providecommand \selectlanguage [0]{\@gobble}%
\providecommand \bibinfo [0]{\@secondoftwo}%
\providecommand \bibfield [0]{\@secondoftwo}%
\providecommand \translation [1]{[#1]}%
\providecommand \BibitemOpen[0]{}%
\providecommand \bibitemStop [0]{}%
\providecommand \bibitemNoStop [0]{.\EOS\space}%
\providecommand \EOS [0]{\spacefactor3000\relax}%
\providecommand \BibitemShut [1]{\csname bibitem#1\endcsname}%
%</preamble>
\bibitem{Eriksen2004}%
  \BibitemOpen
  \bibfield{author}{%
  \bibinfo {author} {\bibfnamefont{H.~K.}\ \bibnamefont{{Eriksen}}}, \bibinfo
  {author} {\bibfnamefont{F.~K.}\ \bibnamefont{{Hansen}}}, \bibinfo {author}
  {\bibfnamefont{A.~J.}\ \bibnamefont{{Banday}}}, \bibinfo {author}
  {\bibfnamefont{K.~M.}\ \bibnamefont{{G{\'o}rski}}},\ and\ \bibinfo {author}
  {\bibfnamefont{P.~B.}\ \bibnamefont{{Lilje}}},\ }%
  \bibfield{journal}{%
  \Doi{10.1086/382267}{\bibinfo {journal} {Astrophys.J.}}\ }%
  \textbf{\bibinfo {volume} {605}},\ \bibinfo {pages} {14} (\bibinfo {month}
  {Apr.}\ \bibinfo {year} {2004}),\
  \Eprint{http://arxiv.org/abs/astro-ph/0307507}{astro-ph/0307507}%
  \bibAnnoteFile{NoStop}{Eriksen2004}%
\bibitem{Eriksen2007}%
  \BibitemOpen
  \bibfield{author}{%
  \bibinfo {author} {\bibfnamefont{H.~K.}\ \bibnamefont{{Eriksen}}}, \bibinfo
  {author} {\bibfnamefont{A.~J.}\ \bibnamefont{{Banday}}}, \bibinfo {author}
  {\bibfnamefont{K.~M.}\ \bibnamefont{{G{\'o}rski}}}, \bibinfo {author}
  {\bibfnamefont{F.~K.}\ \bibnamefont{{Hansen}}},\ and\ \bibinfo {author}
  {\bibfnamefont{P.~B.}\ \bibnamefont{{Lilje}}},\ }%
  \bibfield{journal}{%
  \Doi{10.1086/518091}{\bibinfo {journal} {Astrophys.J.}}\ }%
  \textbf{\bibinfo {volume} {660}},\ \bibinfo {pages} {L81} (\bibinfo {month}
  {May}\ \bibinfo {year} {2007}),\
  \Eprint{http://arxiv.org/abs/astro-ph/0701089}{astro-ph/0701089}%
  \bibAnnoteFile{NoStop}{Eriksen2007}%
\bibitem{Erickcek2008}%
  \BibitemOpen
  \bibfield{author}{%
  \bibinfo {author} {\bibfnamefont{A.~L.}\ \bibnamefont{{Erickcek}}}, \bibinfo
  {author} {\bibfnamefont{M.}~\bibnamefont{{Kamionkowski}}},\ and\ \bibinfo
  {author} {\bibfnamefont{S.~M.}\ \bibnamefont{{Carroll}}},\ }%
  \bibfield{journal}{%
  \Doi{10.1103/PhysRevD.78.123520}{\bibinfo {journal} {Phys. Rev. D}}\ }%
  \textbf{\bibinfo {volume} {78}},\ \bibinfo {eid} {123520} (\bibinfo {month}
  {Dec.}\ \bibinfo {year} {2008}),\
  \Eprint{http://arxiv.org/abs/0806.0377}{arXiv:0806.0377}%
  \bibAnnoteFile{NoStop}{Erickcek2008}%
\bibitem{Hansen2009}%
  \BibitemOpen
  \bibfield{author}{%
  \bibinfo {author} {\bibfnamefont{F.~K.}\ \bibnamefont{{Hansen}}}, \bibinfo
  {author} {\bibfnamefont{A.~J.}\ \bibnamefont{{Banday}}}, \bibinfo {author}
  {\bibfnamefont{K.~M.}\ \bibnamefont{{G{\'o}rski}}}, \bibinfo {author}
  {\bibfnamefont{H.~K.}\ \bibnamefont{{Eriksen}}},\ and\ \bibinfo {author}
  {\bibfnamefont{P.~B.}\ \bibnamefont{{Lilje}}},\ }%
  \bibfield{journal}{%
  \Doi{10.1088/0004-637X/704/2/1448}{\bibinfo {journal} {Astrophys.J.}}\ }%
  \textbf{\bibinfo {volume} {704}},\ \bibinfo {pages} {1448} (\bibinfo {month}
  {Oct.}\ \bibinfo {year} {2009}),\
  \Eprint{http://arxiv.org/abs/0812.3795}{arXiv:0812.3795}%
  \bibAnnoteFile{NoStop}{Hansen2009}%
\bibitem{Hoftuft2009}%
  \BibitemOpen
  \bibfield{author}{%
  \bibinfo {author} {\bibfnamefont{J.}~\bibnamefont{{Hoftuft}}}, \bibinfo
  {author} {\bibfnamefont{H.~K.}\ \bibnamefont{{Eriksen}}}, \bibinfo {author}
  {\bibfnamefont{A.~J.}\ \bibnamefont{{Banday}}}, \bibinfo {author}
  {\bibfnamefont{K.~M.}\ \bibnamefont{{G{\'o}rski}}}, \bibinfo {author}
  {\bibfnamefont{F.~K.}\ \bibnamefont{{Hansen}}},\ and\ \bibinfo {author}
  {\bibfnamefont{P.~B.}\ \bibnamefont{{Lilje}}},\ }%
  \bibfield{journal}{%
  \Doi{10.1088/0004-637X/699/2/985}{\bibinfo {journal} {Astrophys.J.}}\ }%
  \textbf{\bibinfo {volume} {699}},\ \bibinfo {pages} {985} (\bibinfo {month}
  {Jul.}\ \bibinfo {year} {2009}),\
  \Eprint{http://arxiv.org/abs/0903.1229}{arXiv:0903.1229 [astro-ph.CO]}%
  \bibAnnoteFile{NoStop}{Hoftuft2009}%
\bibitem{Paci2013}%
  \BibitemOpen
  \bibfield{author}{%
  \bibinfo {author} {\bibfnamefont{F.}~\bibnamefont{{Paci}}}, \bibinfo {author}
  {\bibfnamefont{A.}~\bibnamefont{{Gruppuso}}}, \bibinfo {author}
  {\bibfnamefont{F.}~\bibnamefont{{Finelli}}}, \bibinfo {author}
  {\bibfnamefont{A.}~\bibnamefont{{De Rosa}}}, \bibinfo {author}
  {\bibfnamefont{N.}~\bibnamefont{{Mandolesi}}},\ and\ \bibinfo {author}
  {\bibfnamefont{P.}~\bibnamefont{{Natoli}}},\ }%
  \bibfield{journal}{%
  \Doi{10.1093/mnras/stt1219}{\bibinfo {journal} {MNRAS}}\ }%
  \textbf{\bibinfo {volume} {434}},\ \bibinfo {pages} {3071} (\bibinfo {month}
  {Oct.}\ \bibinfo {year} {2013}),\
  \Eprint{http://arxiv.org/abs/1301.5195}{arXiv:1301.5195}%
  \bibAnnoteFile{NoStop}{Paci2013}%
\bibitem{Planck2013a}%
  \BibitemOpen
  \bibfield{author}{%
  \bibinfo {author} {\bibnamefont{{Planck Collaboration}}}, \bibinfo {author}
  {\bibfnamefont{P.~A.~R.}\ \bibnamefont{{Ade}}}, \bibinfo {author}
  {\bibfnamefont{N.}~\bibnamefont{{Aghanim}}}, \bibinfo {author}
  {\bibfnamefont{C.}~\bibnamefont{{Armitage-Caplan}}}, \bibinfo {author}
  {\bibfnamefont{M.}~\bibnamefont{{Arnaud}}}, \bibinfo {author}
  {\bibfnamefont{M.}~\bibnamefont{{Ashdown}}}, \bibinfo {author}
  {\bibfnamefont{F.}~\bibnamefont{{Atrio-Barandela}}}, \bibinfo {author}
  {\bibfnamefont{J.}~\bibnamefont{{Aumont}}}, \bibinfo {author}
  {\bibfnamefont{C.}~\bibnamefont{{Baccigalupi}}}, \bibinfo {author}
  {\bibfnamefont{A.~J.}\ \bibnamefont{{Banday}}},\ and\ \bibinfo {author}
  {\bibnamefont{et~al.}},\ }%
  \bibfield{journal}{%
  \Doi{10.1051/0004-6361/201321534}{\bibinfo {journal} {Astron.Astrophys.}}\ }%
  \textbf{\bibinfo {volume} {571}},\ \bibinfo {eid} {A23} (\bibinfo {month}
  {Nov.}\ \bibinfo {year} {2014}),\
  \Eprint{http://arxiv.org/abs/1303.5083}{arXiv:1303.5083}%
  \bibAnnoteFile{NoStop}{Planck2013a}%
\bibitem{Akrami2014}%
  \BibitemOpen
  \bibfield{author}{%
  \bibinfo {author} {\bibfnamefont{Y.}~\bibnamefont{{Akrami}}}, \bibinfo
  {author} {\bibfnamefont{Y.}~\bibnamefont{{Fantaye}}}, \bibinfo {author}
  {\bibfnamefont{A.}~\bibnamefont{{Shafieloo}}}, \bibinfo {author}
  {\bibfnamefont{H.~K.}\ \bibnamefont{{Eriksen}}}, \bibinfo {author}
  {\bibfnamefont{F.~K.}\ \bibnamefont{{Hansen}}}, \bibinfo {author}
  {\bibfnamefont{A.~J.}\ \bibnamefont{{Banday}}},\ and\ \bibinfo {author}
  {\bibfnamefont{K.~M.}\ \bibnamefont{{G{\'o}rski}}},\ }%
  \bibfield{journal}{%
  \Doi{10.1088/2041-8205/784/2/L42}{\bibinfo {journal} {Astrophys.J.}}\ }%
  \textbf{\bibinfo {volume} {784}},\ \bibinfo {eid} {L42} (\bibinfo {month}
  {Apr.}\ \bibinfo {year} {2014}),\
  \Eprint{http://arxiv.org/abs/1402.0870}{arXiv:1402.0870}%
  \bibAnnoteFile{NoStop}{Akrami2014}%
\bibitem{Gordon2005}%
  \BibitemOpen
  \bibfield{author}{%
  \bibinfo {author} {\bibfnamefont{C.}~\bibnamefont{{Gordon}}}, \bibinfo
  {author} {\bibfnamefont{W.}~\bibnamefont{{Hu}}}, \bibinfo {author}
  {\bibfnamefont{D.}~\bibnamefont{{Huterer}}},\ and\ \bibinfo {author}
  {\bibfnamefont{T.}~\bibnamefont{{Crawford}}},\ }%
  \bibfield{journal}{%
  \Doi{10.1103/PhysRevD.72.103002}{\bibinfo {journal} {Phys. Rev. D}}\ }%
  \textbf{\bibinfo {volume} {72}},\ \bibinfo {eid} {103002} (\bibinfo {month}
  {Nov.}\ \bibinfo {year} {2005}),\
  \Eprint{http://arxiv.org/abs/astro-ph/0509301}{astro-ph/0509301}%
  \bibAnnoteFile{NoStop}{Gordon2005}%
\bibitem{Gordon2007}%
  \BibitemOpen
  \bibfield{author}{%
  \bibinfo {author} {\bibfnamefont{C.}~\bibnamefont{{Gordon}}},\ }%
  \bibfield{journal}{%
  \Doi{10.1086/510511}{\bibinfo {journal} {Astrophys.J.}}\ }%
  \textbf{\bibinfo {volume} {656}},\ \bibinfo {pages} {636} (\bibinfo {month}
  {Feb.}\ \bibinfo {year} {2007}),\
  \Eprint{http://arxiv.org/abs/astro-ph/0607423}{astro-ph/0607423}%
  \bibAnnoteFile{NoStop}{Gordon2007}%
\bibitem{Prunet2005}%
  \BibitemOpen
  \bibfield{author}{%
  \bibinfo {author} {\bibfnamefont{S.}~\bibnamefont{{Prunet}}}, \bibinfo
  {author} {\bibfnamefont{J.-P.}\ \bibnamefont{{Uzan}}}, \bibinfo {author}
  {\bibfnamefont{F.}~\bibnamefont{{Bernardeau}}},\ and\ \bibinfo {author}
  {\bibfnamefont{T.}~\bibnamefont{{Brunier}}},\ }%
  \bibfield{journal}{%
  \Doi{10.1103/PhysRevD.71.083508}{\bibinfo {journal} {Phys. Rev. D}}\ }%
  \textbf{\bibinfo {volume} {71}},\ \bibinfo {eid} {083508} (\bibinfo {month}
  {Apr.}\ \bibinfo {year} {2005}),\
  \Eprint{http://arxiv.org/abs/astro-ph/0406364}{astro-ph/0406364}%
  \bibAnnoteFile{NoStop}{Prunet2005}%
\bibitem{Bennett2011}%
  \BibitemOpen
  \bibfield{author}{%
  \bibinfo {author} {\bibfnamefont{C.~L.}\ \bibnamefont{{Bennett}}}, \bibinfo
  {author} {\bibfnamefont{R.~S.}\ \bibnamefont{{Hill}}}, \bibinfo {author}
  {\bibfnamefont{G.}~\bibnamefont{{Hinshaw}}}, \bibinfo {author}
  {\bibfnamefont{D.}~\bibnamefont{{Larson}}}, \bibinfo {author}
  {\bibfnamefont{K.~M.}\ \bibnamefont{{Smith}}}, \bibinfo {author}
  {\bibfnamefont{J.}~\bibnamefont{{Dunkley}}}, \bibinfo {author}
  {\bibfnamefont{B.}~\bibnamefont{{Gold}}}, \bibinfo {author}
  {\bibfnamefont{M.}~\bibnamefont{{Halpern}}}, \bibinfo {author}
  {\bibfnamefont{N.}~\bibnamefont{{Jarosik}}}, \bibinfo {author}
  {\bibfnamefont{A.}~\bibnamefont{{Kogut}}}, \bibinfo {author}
  {\bibfnamefont{E.}~\bibnamefont{{Komatsu}}}, \bibinfo {author}
  {\bibfnamefont{M.}~\bibnamefont{{Limon}}}, \bibinfo {author}
  {\bibfnamefont{S.~S.}\ \bibnamefont{{Meyer}}}, \bibinfo {author}
  {\bibfnamefont{M.~R.}\ \bibnamefont{{Nolta}}}, \bibinfo {author}
  {\bibfnamefont{N.}~\bibnamefont{{Odegard}}}, \bibinfo {author}
  {\bibfnamefont{L.}~\bibnamefont{{Page}}}, \bibinfo {author}
  {\bibfnamefont{D.~N.}\ \bibnamefont{{Spergel}}}, \bibinfo {author}
  {\bibfnamefont{G.~S.}\ \bibnamefont{{Tucker}}}, \bibinfo {author}
  {\bibfnamefont{J.~L.}\ \bibnamefont{{Weiland}}}, \bibinfo {author}
  {\bibfnamefont{E.}~\bibnamefont{{Wollack}}},\ and\ \bibinfo {author}
  {\bibfnamefont{E.~L.}\ \bibnamefont{{Wright}}},\ }%
  \bibfield{journal}{%
  \Doi{10.1088/0067-0049/192/2/17}{\bibinfo {journal} {Astrophys.J.Suppl.}}\ }%
  \textbf{\bibinfo {volume} {192}},\ \bibinfo {eid} {17} (\bibinfo {month}
  {Feb.}\ \bibinfo {year} {2011}),\
  \Eprint{http://arxiv.org/abs/1001.4758}{arXiv:1001.4758 [astro-ph.CO]}%
  \bibAnnoteFile{NoStop}{Bennett2011}%
\bibitem{Rath:2013}%
  \BibitemOpen
  \bibfield{author}{%
  \bibinfo {author} {\bibfnamefont{P.~K.}\ \bibnamefont{Rath}}\ and\ \bibinfo
  {author} {\bibfnamefont{P.}~\bibnamefont{Jain}},\ }%
  \bibfield{journal}{%
  \Doi{10.1088/1475-7516/2013/12/014}{\bibinfo {journal} {JCAP}}\ }%
  \textbf{\bibinfo {volume} {1312}},\ \bibinfo {pages} {014} (\bibinfo {year}
  {2013}),\ \Eprint{http://arxiv.org/abs/1308.0924}{arXiv:1308.0924
  [astro-ph.CO]}%
  \bibAnnoteFile{NoStop}{Rath:2013}%
%%CITATION = ARXIV:1308.0924;%%
\bibitem{Namjoo:2014pqa}%
  \BibitemOpen
  \bibfield{author}{%
  \bibinfo {author} {\bibfnamefont{M.}~\bibnamefont{Namjoo}}, \bibinfo {author}
  {\bibfnamefont{A.}~\bibnamefont{Abolhasani}}, \bibinfo {author}
  {\bibfnamefont{H.}~\bibnamefont{Assadullahi}}, \bibinfo {author}
  {\bibfnamefont{S.}~\bibnamefont{Baghram}}, \bibinfo {author}
  {\bibfnamefont{H.}~\bibnamefont{Firouzjahi}}, \emph{et~al.},\ }%
  \bibfield{journal}{%
  \Doi{10.1088/1475-7516/2015/05/015}{\bibinfo {journal} {JCAP}}\ }%
  \textbf{\bibinfo {volume} {1505}},\ \bibinfo {pages} {015} (\bibinfo {year}
  {2015}),\ \Eprint{http://arxiv.org/abs/1411.5312}{arXiv:1411.5312
  [astro-ph.CO]}%
  \bibAnnoteFile{NoStop}{Namjoo:2014pqa}%
%%CITATION = ARXIV:1411.5312;%%
\bibitem{Kothari:2015}%
  \BibitemOpen
  \bibfield{author}{%
  \bibinfo {author} {\bibfnamefont{R.}~\bibnamefont{Kothari}}, \bibinfo
  {author} {\bibfnamefont{S.}~\bibnamefont{Ghosh}}, \bibinfo {author}
  {\bibfnamefont{P.~K.}\ \bibnamefont{Rath}}, \bibinfo {author}
  {\bibfnamefont{G.}~\bibnamefont{Kashyap}},\ and\ \bibinfo {author}
  {\bibfnamefont{P.}~\bibnamefont{Jain}}}%
   (\bibinfo {year} {2015}),\
  \Eprint{http://arxiv.org/abs/1503.08997}{arXiv:1503.08997 [astro-ph.CO]}%
  \bibAnnoteFile{NoStop}{Kothari:2015}%
%%CITATION = ARXIV:1503.08997;%%
\bibitem{Zaldarriaga:1996}%
  \BibitemOpen
  \bibfield{author}{%
  \bibinfo {author} {\bibfnamefont{M.}~\bibnamefont{Zaldarriaga}}\ and\
  \bibinfo {author} {\bibfnamefont{U.}~\bibnamefont{Seljak}},\ }%
  \bibfield{journal}{%
  \Doi{10.1103/PhysRevD.55.1830}{\bibinfo {journal} {Phys.Rev.}}\ }%
  \textbf{\bibinfo {volume} {D55}},\ \bibinfo {pages} {1830} (\bibinfo {year}
  {1997}),\
  \Eprint{http://arxiv.org/abs/astro-ph/9609170}{arXiv:astro-ph/9609170
  [astro-ph]}%
  \bibAnnoteFile{NoStop}{Zaldarriaga:1996}%
%%CITATION = ASTRO-PH/9609170;%%
\bibitem{Kamionkowski:1996}%
  \BibitemOpen
  \bibfield{author}{%
  \bibinfo {author} {\bibfnamefont{M.}~\bibnamefont{Kamionkowski}}, \bibinfo
  {author} {\bibfnamefont{A.}~\bibnamefont{Kosowsky}},\ and\ \bibinfo {author}
  {\bibfnamefont{A.}~\bibnamefont{Stebbins}},\ }%
  \bibfield{journal}{%
  \Doi{10.1103/PhysRevD.55.7368}{\bibinfo {journal} {Phys.Rev.}}\ }%
  \textbf{\bibinfo {volume} {D55}},\ \bibinfo {pages} {7368} (\bibinfo {year}
  {1997}),\
  \Eprint{http://arxiv.org/abs/astro-ph/9611125}{arXiv:astro-ph/9611125
  [astro-ph]}%
  \bibAnnoteFile{NoStop}{Kamionkowski:1996}%
%%CITATION = ASTRO-PH/9611125;%%
\bibitem{Newman:1966}%
  \BibitemOpen
  \bibfield{author}{%
  \bibinfo {author} {\bibfnamefont{E.~T.}\ \bibnamefont{Newman}}\ and\ \bibinfo
  {author} {\bibfnamefont{R.}~\bibnamefont{Penrose}},\ }%
  \bibfield{journal}{%
  \Doi{http://dx.doi.org/10.1063/1.1931221}{\bibinfo {journal} {Journal of
  Mathematical Physics}}\ }%
  \textbf{\bibinfo {volume} {7}},\ \bibinfo {pages} {863} (\bibinfo {year}
  {1966})%
  \bibAnnoteFile{NoStop}{Newman:1966}%
\bibitem{Goldberg:1967}%
  \BibitemOpen
  \bibfield{author}{%
  \bibinfo {author} {\bibfnamefont{J.~N.}\ \bibnamefont{{Goldberg}}}, \bibinfo
  {author} {\bibfnamefont{A.~J.}\ \bibnamefont{{Macfarlane}}}, \bibinfo
  {author} {\bibfnamefont{E.~T.}\ \bibnamefont{{Newman}}}, \bibinfo {author}
  {\bibfnamefont{F.}~\bibnamefont{{Rohrlich}}},\ and\ \bibinfo {author}
  {\bibfnamefont{E.~C.~G.}\ \bibnamefont{{Sudarshan}}},\ }%
  \bibfield{journal}{%
  \Doi{10.1063/1.1705135}{\bibinfo {journal} {Journal of Mathematical
  Physics}}\ }%
  \textbf{\bibinfo {volume} {8}},\ \bibinfo {pages} {2155} (\bibinfo {month}
  {Nov.}\ \bibinfo {year} {1967})%
  \bibAnnoteFile{NoStop}{Goldberg:1967}%
\bibitem{1972hmfw.book.....A}%
  \BibitemOpen
  \bibfield{author}{%
  \bibinfo {author} {\bibfnamefont{M.}~\bibnamefont{{Abramowitz}}}\ and\
  \bibinfo {author} {\bibfnamefont{I.~A.}\ \bibnamefont{{Stegun}}},\ }%
  \emph{\bibinfo {title} {Handbook of Mathematical Functions, New York: Dover,
  1972}}\ (\bibinfo {year} {1972})%
  \bibAnnoteFile{NoStop}{1972hmfw.book.....A}%
\bibitem{Gallier:2011}%
  \BibitemOpen
  \bibfield{author}{%
  \bibinfo {author} {\bibfnamefont{J.}~\bibnamefont{{Gallier}}},\ }%
  \emph{\bibinfo {title} {Discrete Mathematics, Springer, 2011}}\ (\bibinfo
  {year} {2011})%
  \bibAnnoteFile{NoStop}{Gallier:2011}%
\bibitem{Rath:2014}%
  \BibitemOpen
  \bibfield{author}{%
  \bibinfo {author} {\bibfnamefont{P.~K.}\ \bibnamefont{Rath}}, \bibinfo
  {author} {\bibfnamefont{P.~K.}\ \bibnamefont{Aluri}},\ and\ \bibinfo {author}
  {\bibfnamefont{P.}~\bibnamefont{Jain}},\ }%
  \bibfield{journal}{%
  \Doi{10.1103/PhysRevD.91.023515}{\bibinfo {journal} {Phys.Rev.}}\ }%
  \textbf{\bibinfo {volume} {D91}},\ \bibinfo {pages} {023515} (\bibinfo {year}
  {2015}),\ \Eprint{http://arxiv.org/abs/1403.2567}{arXiv:1403.2567
  [astro-ph.CO]}%
  \bibAnnoteFile{NoStop}{Rath:2014}%
%%CITATION = ARXIV:1403.2567;%%
\bibitem{Gorski:2004}%
  \BibitemOpen
  \bibfield{author}{%
  \bibinfo {author} {\bibfnamefont{K.}~\bibnamefont{Gorski}}, \bibinfo {author}
  {\bibfnamefont{E.}~\bibnamefont{Hivon}}, \bibinfo {author}
  {\bibfnamefont{A.}~\bibnamefont{Banday}}, \bibinfo {author}
  {\bibfnamefont{B.}~\bibnamefont{Wandelt}}, \bibinfo {author}
  {\bibfnamefont{F.}~\bibnamefont{Hansen}}, \emph{et~al.},\ }%
  \bibfield{journal}{%
  \Doi{10.1086/427976}{\bibinfo {journal} {Astrophys.J.}}\ }%
  \textbf{\bibinfo {volume} {622}},\ \bibinfo {pages} {759} (\bibinfo {year}
  {2005}),\
  \Eprint{http://arxiv.org/abs/astro-ph/0409513}{arXiv:astro-ph/0409513
  [astro-ph]}%
  \bibAnnoteFile{NoStop}{Gorski:2004}%
%%CITATION = ASTRO-PH/0409513;%%
\bibitem{Lewis:1999}%
  \BibitemOpen
  \bibfield{author}{%
  \bibinfo {author} {\bibfnamefont{A.}~\bibnamefont{Lewis}}, \bibinfo {author}
  {\bibfnamefont{A.}~\bibnamefont{Challinor}},\ and\ \bibinfo {author}
  {\bibfnamefont{A.}~\bibnamefont{Lasenby}},\ }%
  \bibfield{journal}{%
  \Doi{10.1086/309179}{\bibinfo {journal} {Astrophys.J.}}\ }%
  \textbf{\bibinfo {volume} {538}},\ \bibinfo {pages} {473} (\bibinfo {year}
  {2000}),\
  \Eprint{http://arxiv.org/abs/astro-ph/9911177}{arXiv:astro-ph/9911177
  [astro-ph]}%
  \bibAnnoteFile{NoStop}{Lewis:1999}%
%%CITATION = ASTRO-PH/9911177;%%
\bibitem{Ade:2015}%
  \BibitemOpen
  \bibfield{author}{%
  \bibinfo {author} {\bibfnamefont{P.}~\bibnamefont{Ade}} \emph{et~al.}
  (\bibinfo {collaboration} {Planck})}%
   (\bibinfo {year} {2015}),\
  \Eprint{http://arxiv.org/abs/1502.01589}{arXiv:1502.01589 [astro-ph.CO]}%
  \bibAnnoteFile{NoStop}{Ade:2015}%
%%CITATION = ARXIV:1502.01589;%%
\bibitem{Ade:2015XVI}%
  \BibitemOpen
  \bibfield{author}{%
  \bibinfo {author} {\bibfnamefont{P.}~\bibnamefont{Ade}} \emph{et~al.}
  (\bibinfo {collaboration} {Planck})}%
   (\bibinfo {year} {2015}),\
  \Eprint{http://arxiv.org/abs/1506.07135}{arXiv:1506.07135 [astro-ph.CO]}%
  \bibAnnoteFile{NoStop}{Ade:2015XVI}%
%%CITATION = ARXIV:1506.07135;%%
\bibitem{Aiola:2015}%
  \BibitemOpen
  \bibfield{author}{%
  \bibinfo {author} {\bibfnamefont{S.}~\bibnamefont{{Aiola}}}, \bibinfo
  {author} {\bibfnamefont{B.}~\bibnamefont{{Wang}}}, \bibinfo {author}
  {\bibfnamefont{A.}~\bibnamefont{{Kosowsky}}}, \bibinfo {author}
  {\bibfnamefont{T.}~\bibnamefont{{Kahniashvili}}},\ and\ \bibinfo {author}
  {\bibfnamefont{H.}~\bibnamefont{{Firouzjahi}}},\ }%
  \bibfield{journal}{%
  \bibinfo {journal} {ArXiv e-prints}}%
   (\bibinfo {month} {Jun.}\ \bibinfo {year} {2015}),\
  \Eprint{http://arxiv.org/abs/1506.04405}{arXiv:1506.04405}%
  \bibAnnoteFile{NoStop}{Aiola:2015}%
\bibitem{Donoghue:2005}%
  \BibitemOpen
  \bibfield{author}{%
  \bibinfo {author} {\bibfnamefont{E.~P.}\ \bibnamefont{{Donoghue}}}\ and\
  \bibinfo {author} {\bibfnamefont{J.~F.}\ \bibnamefont{{Donoghue}}},\ }%
  \bibfield{journal}{%
  \Doi{10.1103/PhysRevD.71.043002}{\bibinfo {journal} {Phys. Rev. D}}\ }%
  \textbf{\bibinfo {volume} {71}},\ \bibinfo {eid} {043002} (\bibinfo {month}
  {Feb.}\ \bibinfo {year} {2005}),\
  \Eprint{http://arxiv.org/abs/astro-ph/0411237}{astro-ph/0411237}%
  \bibAnnoteFile{NoStop}{Donoghue:2005}%
\bibitem{Hanson:2009}%
  \BibitemOpen
  \bibfield{author}{%
  \bibinfo {author} {\bibfnamefont{D.}~\bibnamefont{{Hanson}}}\ and\ \bibinfo
  {author} {\bibfnamefont{A.}~\bibnamefont{{Lewis}}},\ }%
  \bibfield{journal}{%
  \Doi{10.1103/PhysRevD.80.063004}{\bibinfo {journal} {Phys. Rev. D}}\ }%
  \textbf{\bibinfo {volume} {80}},\ \bibinfo {eid} {063004} (\bibinfo {month}
  {Sep.}\ \bibinfo {year} {2009}),\
  \Eprint{http://arxiv.org/abs/0908.0963}{arXiv:0908.0963 [astro-ph.CO]}%
  \bibAnnoteFile{NoStop}{Hanson:2009}%
\bibitem{Quartin:2014}%
  \BibitemOpen
  \bibfield{author}{%
  \bibinfo {author} {\bibfnamefont{M.}~\bibnamefont{{Quartin}}}\ and\ \bibinfo
  {author} {\bibfnamefont{A.}~\bibnamefont{{Notari}}},\ }%
  \bibfield{journal}{%
  \Doi{10.1088/1475-7516/2015/01/008}{\bibinfo {journal} {JCAP}}\ }%
  \textbf{\bibinfo {volume} {1}},\ \bibinfo {eid} {008} (\bibinfo {month}
  {Jan.}\ \bibinfo {year} {2015}),\
  \Eprint{http://arxiv.org/abs/1408.5792}{arXiv:1408.5792}%
  \bibAnnoteFile{NoStop}{Quartin:2014}%
\bibitem{Ralston:2004}%
  \BibitemOpen
  \bibfield{author}{%
  \bibinfo {author} {\bibfnamefont{J.~P.}\ \bibnamefont{Ralston}}\ and\
  \bibinfo {author} {\bibfnamefont{P.}~\bibnamefont{Jain}},\ }%
  \bibfield{journal}{%
  \Doi{10.1142/S0218271804005948}{\bibinfo {journal} {IJMPD}}\ }%
  \textbf{\bibinfo {volume} {13}},\ \bibinfo {pages} {1857} (\bibinfo {year}
  {2004}),\
  \Eprint{http://arxiv.org/abs/astro-ph/0311430}{arXiv:astro-ph/0311430
  [astro-ph]}%
  \bibAnnoteFile{NoStop}{Ralston:2004}%
%%CITATION = ASTRO-PH/0311430;%%
\bibitem{Schwarz:2004}%
  \BibitemOpen
  \bibfield{author}{%
  \bibinfo {author} {\bibfnamefont{D.~J.}\ \bibnamefont{Schwarz}}, \bibinfo
  {author} {\bibfnamefont{G.~D.}\ \bibnamefont{Starkman}}, \bibinfo {author}
  {\bibfnamefont{D.}~\bibnamefont{Huterer}},\ and\ \bibinfo {author}
  {\bibfnamefont{C.~J.}\ \bibnamefont{Copi}},\ }%
  \bibfield{journal}{%
  \Doi{10.1103/PhysRevLett.93.221301}{\bibinfo {journal} {Phys. Rev. Lett.}}\
  }%
  \textbf{\bibinfo {volume} {93}},\ \bibinfo {pages} {221301} (\bibinfo {month}
  {Nov}\ \bibinfo {year} {2004})%
  \bibAnnoteFile{NoStop}{Schwarz:2004}%
\bibitem{Jain:1998}%
  \BibitemOpen
  \bibfield{author}{%
  \bibinfo {author} {\bibfnamefont{P.}~\bibnamefont{Jain}}\ and\ \bibinfo
  {author} {\bibfnamefont{J.~P.}\ \bibnamefont{Ralston}},\ }%
  \bibfield{journal}{%
  \Doi{10.1142/S0217732399000481}{\bibinfo {journal} {Mod. Phys. Lett. A}}\ }%
  \textbf{\bibinfo {volume} {14}},\ \bibinfo {pages} {417} (\bibinfo {year}
  {1999})%
  \bibAnnoteFile{NoStop}{Jain:1998}%
\bibitem{deOliveiraCosta:2003}%
  \BibitemOpen
  \bibfield{author}{%
  \bibinfo {author} {\bibfnamefont{A.}~\bibnamefont{de~Oliveira-Costa}},
  \bibinfo {author} {\bibfnamefont{M.}~\bibnamefont{Tegmark}}, \bibinfo
  {author} {\bibfnamefont{M.}~\bibnamefont{Zaldarriaga}},\ and\ \bibinfo
  {author} {\bibfnamefont{A.}~\bibnamefont{Hamilton}},\ }%
  \bibfield{journal}{%
  \Doi{10.1103/PhysRevD.69.063516}{\bibinfo {journal} {Phys.Rev.}}\ }%
  \textbf{\bibinfo {volume} {D69}},\ \bibinfo {pages} {063516} (\bibinfo {year}
  {2004}),\
  \Eprint{http://arxiv.org/abs/astro-ph/0307282}{arXiv:astro-ph/0307282
  [astro-ph]}%
  \bibAnnoteFile{NoStop}{deOliveiraCosta:2003}%
%%CITATION = ASTRO-PH/0307282;%%
\bibitem{Singal:2011}%
  \BibitemOpen
  \bibfield{author}{%
  \bibinfo {author} {\bibfnamefont{A.~K.}\ \bibnamefont{Singal}},\ }%
  \bibfield{journal}{%
  \bibinfo {journal} {ApJL}\ }%
  \textbf{\bibinfo {volume} {742}},\ \bibinfo {pages} {L23} (\bibinfo {year}
  {2011}),\ \Eprint{http://arxiv.org/abs/1110.6260}{arXiv:1110.6260
  [astro-ph.CO]}%
  \bibAnnoteFile{NoStop}{Singal:2011}%
%%CITATION = ARXIV:1110.6260;%%
\bibitem{Gibelyou:2012}%
  \BibitemOpen
  \bibfield{author}{%
  \bibinfo {author} {\bibfnamefont{C.}~\bibnamefont{Gibelyou}}\ and\ \bibinfo
  {author} {\bibfnamefont{D.}~\bibnamefont{Huterer}},\ }%
  \bibfield{journal}{%
  \Doi{10.1111/j.1365-2966.2012.22032.x}{\bibinfo {journal} {MNRAS}}\ }%
  \textbf{\bibinfo {volume} {427}},\ \bibinfo {pages} {1994} (\bibinfo {year}
  {2012}),\ \Eprint{http://arxiv.org/abs/1205.6476}{arXiv:1205.6476
  [astro-ph.CO]}%
  \bibAnnoteFile{NoStop}{Gibelyou:2012}%
%%CITATION = ARXIV:1205.6476;%%
\bibitem{Rubart:2013}%
  \BibitemOpen
  \bibfield{author}{%
  \bibinfo {author} {\bibfnamefont{M.}~\bibnamefont{Rubart}}\ and\ \bibinfo
  {author} {\bibfnamefont{D.~J.}\ \bibnamefont{Schwarz}},\ }%
  \bibfield{journal}{%
  \Doi{10.1051/0004-6361/201321215}{\bibinfo {journal} {A \& A}}\ }%
  \textbf{\bibinfo {volume} {555}},\ \bibinfo {pages} {A117} (\bibinfo {year}
  {2013}),\ \Eprint{http://arxiv.org/abs/1301.5559}{arXiv:1301.5559
  [astro-ph.CO]}%
  \bibAnnoteFile{NoStop}{Rubart:2013}%
%%CITATION = ARXIV:1301.5559;%%
\bibitem{Kothari:2013}%
  \BibitemOpen
  \bibfield{author}{%
  \bibinfo {author} {\bibfnamefont{R.}~\bibnamefont{Kothari}}, \bibinfo
  {author} {\bibfnamefont{A.}~\bibnamefont{Naskar}}, \bibinfo {author}
  {\bibfnamefont{P.}~\bibnamefont{Tiwari}}, \bibinfo {author}
  {\bibfnamefont{S.}~\bibnamefont{Nadkarni-Ghosh}},\ and\ \bibinfo {author}
  {\bibfnamefont{P.}~\bibnamefont{Jain}},\ }%
  \bibfield{journal}{%
  \Doi{10.1016/j.astropartphys.2014.06.004}{\bibinfo {journal} {Astroparticle
  Physics}}\ }%
  \textbf{\bibinfo {volume} {61}},\ \bibinfo {pages} {1} (\bibinfo {month}
  {Feb.}\ \bibinfo {year} {2015}),\
  \Eprint{http://arxiv.org/abs/1307.1947}{arXiv:1307.1947 [astro-ph.CO]}%
  \bibAnnoteFile{NoStop}{Kothari:2013}%
%%CITATION = ARXIV:1307.1947;%%
\bibitem{Tiwari2015}%
  \BibitemOpen
  \bibfield{author}{%
  \bibinfo {author} {\bibfnamefont{P.}~\bibnamefont{{Tiwari}}}\ and\ \bibinfo
  {author} {\bibfnamefont{P.}~\bibnamefont{{Jain}}},\ }%
  \bibfield{journal}{%
  \Doi{10.1093/mnras/stu2535}{\bibinfo {journal} {MNRAS}}\ }%
  \textbf{\bibinfo {volume} {447}},\ \bibinfo {pages} {2658} (\bibinfo {month}
  {Mar.}\ \bibinfo {year} {2015}),\
  \Eprint{http://arxiv.org/abs/1308.3970}{arXiv:1308.3970}%
  \bibAnnoteFile{NoStop}{Tiwari2015}%
\bibitem{Hutsemekers:1998}%
  \BibitemOpen
  \bibfield{author}{%
  \bibinfo {author} {\bibfnamefont{D.}~\bibnamefont{Hutsem\'{e}kers}},\ }%
  \bibfield{journal}{%
  \bibinfo {journal} {Astron.Astrophys.}\ }%
  \textbf{\bibinfo {volume} {332}},\ \bibinfo {pages} {410} (\bibinfo {year}
  {1998})%
  \bibAnnoteFile{NoStop}{Hutsemekers:1998}%
\bibitem{Hutsemekers:2000fv}%
  \BibitemOpen
  \bibfield{author}{%
  \bibinfo {author} {\bibfnamefont{D.}~\bibnamefont{Hutsem\'{e}kers}}\ and\
  \bibinfo {author} {\bibfnamefont{H.}~\bibnamefont{Lamy}},\ }%
  \bibfield{journal}{%
  \Doi{10.1051/0004-6361:20000443}{\bibinfo {journal} {Astron.Astrophys.}}\ }%
  \textbf{\bibinfo {volume} {367}},\ \bibinfo {pages} {381} (\bibinfo {year}
  {2001})%
  \bibAnnoteFile{NoStop}{Hutsemekers:2000fv}%
\bibitem{Jain:2003sg}%
  \BibitemOpen
  \bibfield{author}{%
  \bibinfo {author} {\bibfnamefont{P.}~\bibnamefont{Jain}}, \bibinfo {author}
  {\bibfnamefont{G.}~\bibnamefont{Narain}},\ and\ \bibinfo {author}
  {\bibfnamefont{S.}~\bibnamefont{Sarala}},\ }%
  \bibfield{journal}{%
  \Doi{10.1111/j.1365-2966.2004.07169.x}{\bibinfo {journal} {MNRAS}}\ }%
  \textbf{\bibinfo {volume} {347}},\ \bibinfo {pages} {394} (\bibinfo {year}
  {2004})%
  \bibAnnoteFile{NoStop}{Jain:2003sg}%
\bibitem{Ruiz:1996}%
  \BibitemOpen
  \bibfield{author}{%
  \bibinfo {author} {\bibfnamefont{S.}~\bibnamefont{Ruiz}},\ }%
  \bibfield{journal}{%
  \bibinfo {journal} {The Mathematical Gazette}\ }%
  \textbf{\bibinfo {volume} {80}},\ \bibinfo {pages} {489} (\bibinfo {year}
  {1996})%
  \bibAnnoteFile{NoStop}{Ruiz:1996}%
\bibitem{Gorski:2005}%
  \BibitemOpen
  \bibfield{author}{%
  \bibinfo {author} {\bibfnamefont{K.}~\bibnamefont{Go\'{r}ski}}, \bibinfo
  {author} {\bibfnamefont{E.}~\bibnamefont{Hivon}}, \bibinfo {author}
  {\bibfnamefont{A.}~\bibnamefont{Banday}}, \bibinfo {author}
  {\bibfnamefont{B.}~\bibnamefont{Wandelt}}, \bibinfo {author}
  {\bibfnamefont{F.}~\bibnamefont{Hansen}}, \emph{et~al.},\ }%
  \bibfield{journal}{%
  \Doi{10.1086/427976}{\bibinfo {journal} {ApJ}}\ }%
  \textbf{\bibinfo {volume} {622}},\ \bibinfo {pages} {759} (\bibinfo {year}
  {2005}),\
  \Eprint{http://arxiv.org/abs/astro-ph/0409513}{arXiv:astro-ph/0409513
  [astro-ph]}%
  \bibAnnoteFile{NoStop}{Gorski:2005}%
CITATION = ASTRO-PH/0409513
\end{thebibliography}%
\end{document}